\documentclass[]
{jpp}

\usepackage{natbib}
\usepackage{graphicx}
\usepackage{epstopdf, epsfig}
\usepackage[utf8]{inputenc}
\usepackage[T1]{fontenc}
\usepackage{amsmath}
\usepackage[usenames,dvipsnames]{color}
\usepackage{float}
\usepackage{mathtools}
\usepackage{amsmath,amssymb,amsbsy}
\usepackage{bbm,bm,mathrsfs,yfonts}
\usepackage[capitalize]{cleveref}
\usepackage{xcolor}
\usepackage{microtype}
\usepackage{soul,xcolor}
\setstcolor{red}
\newcommand{\be}{\begin{equation}}
\newcommand{\ee}{\end{equation}}
\newcommand{\lb }{\left<}
\newcommand{\rb }{\right>}
\renewcommand{\l }{\left}
\renewcommand{\r }{\right}

\DeclarePairedDelimiter\floor{\lfloor}{\rfloor}

\shorttitle{A SOL Drift-Kinetic Model at Arbitrary Collisionality}
\shortauthor{R. Jorge, P. Ricci, and N. F. Loureiro}

\title{
A Drift-Kinetic Analytical Model for SOL Plasma Dynamics at Arbitrary Collisionality
}

\author{R. Jorge\aff{1,2}
  \corresp{\email{rogerio.jorge@epfl.ch}},
  P. Ricci\aff{1}
 \and N. F. Loureiro\aff{3}}

\affiliation{
\aff{1}École Polytechnique Fédérale de Lausanne (EPFL), Swiss Plasma Center (SPC), CH-1015
Lausanne, Switzerland
\aff{2}Instituto de Plasmas e Fusão Nuclear, Instituto Superior Técnico, Universidade de Lisboa, 1049-001 Lisboa, Portugal
\aff{3}Plasma Science and Fusion Center, Massachusetts Institute of Technology, Cambridge MA 02139, USA
}

\begin{document}

\maketitle

\begin{abstract}
A drift-kinetic model to describe the plasma dynamics in the scrape-off layer region of tokamak devices at arbitrary collisionality is derived. Our formulation is based on a gyroaveraged Lagrangian description of the charged particle motion, and the corresponding drift-kinetic Boltzmann equation that includes a full Coulomb collision operator. Using a Hermite-Laguerre velocity space decomposition of the gyroaveraged distribution function, a set of equations to evolve the coefficients of the expansion is presented. By evaluating explicitly the moments of the Coulomb collision operator, distribution functions arbitrarily far from equilibrium can be studied at arbitrary collisionalities. A fluid closure in the high-collisionality limit is presented, and the corresponding fluid equations are compared with previously-derived fluid models.
\end{abstract}

\tableofcontents

\section{Introduction}

The success of the magnetic confinement fusion program relies on our ability to predict the dynamics of the plasma in the tokamak scrape-off layer (SOL). In this region, the plasma is turbulent with fluctuation level of order unity \citep{Ritz1987,Wootton1990,Hidalgo2002,LaBombard2005,DIppolito2011}. The fluctuations are characterized by frequencies lower than the ion gyrofrequency \citep{Endler1995,Agostini2011,Carralero2014,Garcia2015}, and the turbulent eddies, which include coherent radial propagation of filamentary structures \citep{DIppolito2002,DIppolito2011,Carreras2005,Serianni2007}, have a radial extension comparable to the time-averaged SOL pressure gradient length $L_p$ \citep{Zweben2007}.

{In recent years, there has been a significant development of first-principles simulations of the SOL dynamics} with both kinetic \citep{Tskhakaya2012} and gyrokinetic \citep{Xu2007, Shi2015, Chang2017,Shi2017} codes.
However, as kinetic simulations {of the SOL and edge regions} remain prohibitive as they still are computationally extremely expensive, the less demanding fluid simulations are the standard of reference \citep{Dudson2009,Tamain2009,Easy2014,Halpern2016a,Madsen2016}. The fluid simulations are usually based on the drift-reduced Braginskii \citep{Braginskii1965,Zeiler1997} or gyrofluid \citep{Ribeiro2008a,Held2016} models to evolve plasma density, fluid velocity and temperature. Fluid models assume that the distribution function is close to a local Maxwellian, and that scale lengths along the magnetic field are longer than the mean free path.
However, kinetic simulations show that the plasma distribution function is far from Maxwellian in the SOL region \citep{Tskhakaya2008,Lonnroth2006,Battaglia2014} and that collisionless effects in the SOL might become important \citep{Batishchev1997}.
This is expected to be particularly true in ITER and other future devices that will be operated in the high confinement mode (H-mode) regime \citep{Martin2008}. In such cases, a transport barrier is formed that creates a steep pressure gradient at the plasma edge. If the pressure gradient exceeds a threshold value, edge-localized modes (ELMs) are destabilized \citep{Leonard2014}, expelling large amounts of heat and particles to the wall. Describing structures with such high temperatures (and therefore low-collisionality) with respect to the background SOL plasma requires therefore a model that allows for the treatment of arbitrary collision frequencies. Higher moments of the distribution function are needed for a proper SOL description \citep{Hazeltine1998}.

Leveraging the development of previous models {\citep{Hammett1993,Beer1996,Sugama2001,Ji2010,Zocco2011,Schekochihin2016,Hatch2016,Parker2016,Hirvijoki2016,Mandell2017}}, we construct here a moment hierarchy to evolve the SOL plasma dynamics. Our model is valid in arbitrary magnetic field geometries and, making use of the full Coulomb collision operator, at arbitrary collision frequencies. The model is derived within a full-F framework, as the amplitude of the background and fluctuating components of the plasma parameters in the SOL have comparable amplitude. We work within the drift approximation \citep{Hinton1976, Cary2009}, which assumes that plasma quantities have typical frequencies that are small compared to the ion gyrofrequency $\Omega_i=e B /m_i$, and their perpendicular spatial scale is small compared to the ion sound Larmor radius $\rho_s=c_s/\Omega_i$, with $c_s^2 = T_e/m_i$, $T_e$ the electron temperature, $B$ the magnitude of the magnetic field, $e$ the electron charge, and $m_i$ the ion mass. Moreover, we consider a Braginskii ordering \citep{Braginskii1965}, where the species flow velocities are comparable to the ion thermal speed, as opposed to the drift ordering introduced by \citet{Mikhailovskii1971} and extended and corrected by \citet{Catto2004}, where flow velocities are comparable to the diamagnetic drift velocities.

More precisely, denoting {$k_\perp \sim |\nabla_\perp \log \phi|\sim |\nabla_\perp \log n|\sim |\nabla_\perp \log T_{e}|$ and $\omega \sim |\partial_t \log \phi| \sim |\partial_t \log n| \sim |\partial_t \log T_e|$}, with $\phi$ the electrostatic potential and $n$ the plasma density, we introduce the ordering parameter $\epsilon$ such that

\begin{equation}
    \epsilon \sim k_\perp \rho_s \sim \omega/\Omega_i \ll 1.
    \label{eq:ordering}
\end{equation}

\noindent {On the other hand, we let $k_\perp L_p \sim 1$ since turbulent eddies {are observed to have} an extension comparable to the scale lengths of the {time}-averaged quantities.}
These assumptions are in agreement with experimental measurements of SOL plasmas \citep{LaBombard2001,Zweben2004,Myra2013,Carralero2014}.
We also order the electron collision frequency $\nu_{ei}$ as

\begin{equation}
	\frac{\nu_{ei}}{ \Omega_i} \sim \epsilon_\nu \lesssim \epsilon,
	\label{eq:orderingnu}
\end{equation}

In addition, the ion collision frequency ${\nu_i =} \nu_{ii}$ is ordered as $\nu_{ii} \lesssim \epsilon^2 \Omega_i$ that, noticing $\nu_i \sim  \sqrt{{m_e}/{m_i}}(T_e/T_i)^{3/2} \nu_e$ {(with $\nu_e = \nu_{ei}$)}, yields
\begin{equation}
    \left(\frac{\epsilon_\nu}{\epsilon^2}\right)^{2/3}\left(\frac{m_e}{m_i}\right)^{1/3}\lesssim\frac{T_i}{T_e}\lesssim 1.
    \label{eq:titebound}
\end{equation}
The ordering in \cref{eq:titebound} can be used to justify applying our model in the cold ion limit, $T_i \ll T_e$, {but allows for $T_i \sim T_e$}.
We note that in the SOL the ratio $T_i/T_e$ is in the range $1 \lesssim T_i/T_e \lesssim 4$ \citep{Kocan2011}. The ion temperature in this range of values is seen to play a negligible role in determining the SOL turbulent dynamics, usually due to a steeper electron temperature profile compared with the ion one, which is usually below the threshold limit of the ion temperature gradient instability \citep{Kwws}.

The ordering in Eqs. (\ref{eq:ordering})-(\ref{eq:titebound}) is justified in a wide variety of experimental conditions. For example, for a typical JET discharge \citep{Erents2000,Liang2007,Xu2009} with the SOL parameters $B_T = 2.5$ T, $T_e \sim T_i \sim 20$ eV, $n_e \simeq 10^{19}$ m$^{-3}$, and $k_\perp \sim 1$ cm$^{-1}$, we obtain $\epsilon_\nu \sim 0.016$ and $\epsilon \sim 0.0182$. For a medium-size tokamak such as TCV \citep{Rossel2012,Nespoli2017}, estimating  $B_T = 1.5$ T, $T_e \sim T_i \sim 40$ eV, $n_e \simeq 6 \times 10^{18}$, and $k_\perp \sim 1$ cm$^{-1}$, we obtain $\epsilon_\nu \sim 6.2 \times 10^{-3}$ and $\epsilon \sim 0.043$. Finally, for small-size tokamaks such as ISTTOK \citep{Silva2011a,Jorge2016}, with $B_T = 0.5$ T, $T_e \sim T_i \sim 20$ eV, $n_e \simeq 0.8 \times 10^{18}$, and $k_\perp \sim 1$ cm$^{-1}$, we obtain $\epsilon_\nu \sim 0.0072$ and $\epsilon \sim 0.091$. Lower values of $\epsilon_\nu$, as in the presence of ELMs where temperatures can reach up to $100$ eV \citep{Pitts2003}, are also included in the ordering considered here.

Following typical SOL experimental measurements (see, e.g. \citet{Zweben2007,Terry2009,Grulke2014}), we order $k_{\parallel} \sim 1/L_B \sim 1/R$, with $L_B$ the background magnetic field spatial gradient scale and $R$ the tokamak major radius, and take $k_\parallel \rho_s \sim \epsilon^3$. This yields

\begin{equation}
    \frac{k_\parallel}{k_\perp} \sim \epsilon^2,
    \label{eq:kpar1}
\end{equation}

\noindent a lower ratio than the ones used in most drift-kinetic and gyrokinetic deductions \citep{Hahm1988a,Hazeltine1968,Abel2013}. The orderings in \cref{eq:orderingnu,eq:kpar1} imply that

\begin{equation}
    k_\parallel \lambda_{mfp} \sim \sqrt{\frac{m_i}{m_e}}\frac{\epsilon^3}{\epsilon_\nu},
    \label{eq:kparlmfp}
\end{equation}
\noindent which includes both the collisional regime $k_{\parallel} \lambda_{mfp} \lesssim 1$, when $\epsilon_\nu \sim \epsilon$, and the collisionless regime $k_{\parallel} \lambda_{mfp} \gg 1$, when $\epsilon_\nu \ll \epsilon$.

\noindent Finally, the plasma parameter $\beta = n T_e/(B^2/2\mu_0)$ is ordered as $\beta \lesssim \epsilon^3$, {implying that our equations describe plasma dynamics in an electrostatic regime}.

Our model describes the evolution of the moments of the drift-kinetic Boltzmann equation at order $\epsilon^2$, taking into account the effect of collisions through a full Coulomb collision operator. The kinetic equation is based on a Lagrangian description of the charged particle motion.
In the SOL, due to the large fluctuations and the short characteristic gradient width $L_\phi \sim L_p$, a strong electric field is present. To properly retain the effect of a non-negligible $\bm E \times \bm B$ drift, $\bm v_E=-\nabla \phi \times \bm B/B^2$ ($\bm E = -\nabla \phi$ in the electrostatic limit employed here), in the equations of motion, we split the perpendicular component of the particle velocity $\bm v_\perp$ into $\bm v_\perp = \bm v_E + \bm v_\perp'$. In the particle Lagrangian, we keep the resulting term $m v_E^2/2$ associated with the $\bm E \times \bm B$ motion of the gyrocenters, as it will be shown to be of the same order of magnitude as the first-order terms in the Lagrangian (see \citet{Krommes2013} for a discussion on the physical interpretation of this term).

In the kinetic equation, we expand the gyroaveraged distribution function into a Hermite-Laguerre basis, and express the moments of the collision operator in a series of products of the expansion coefficients of the distribution function. For like-species collisions, this expansion is based on the work of \citet{Ji2009}, while we make use of the small mass ratio approximation to obtain electron-ion and ion-electron operators that ensure basic conservation properties.
The system is closed by Poisson's equation, involving explicitly the moments of the distribution function, accurate up to order $\epsilon^2$ (we also present a derivation of Poisson's equation that rigorously includes collisional $\epsilon_\nu$ effects).

This paper is organized as follows. \cref{sec:solparticle} derives the equations of motion of a charged particle in the SOL, and the drift-kinetic Boltzmann equation in a conservative form. In \cref{sec:momentexpansion} we expand the gyroaveraged distribution function in a Hermite-Laguerre basis and obtain the guiding-center moments of the collision operator. In \cref{sec:momenthierarchy} we take moments of the drift-kinetic Boltzmann equation, and deduce the moment-hierarchy equations. \cref{sec:poisson} presents the guiding-center Poisson's equation, accurate up to order $\epsilon^2$. Finally, in \cref{sec:fluidmodel}, a fluid model based on the truncation of the Hermite-Laguerre expansion in the high-collisionality regime is presented. The conclusions follow. Appendix \ref{app:tlkpj} presents the transformation between pitch-angle and parallel-perpendicular velocity basis. Appendix \ref{app:collcoef} lists explicitly the moments of the parallel acceleration phase-space conserving term. Appendix \ref{app:poisson} derives Poisson's equation with higher-order collisional effects. Finally, in Appendix \ref{app:cabmoments}, the the lower order guiding-center moments of the collision operator are given in explicit form.

\section{SOL Guiding-Center Model}
\label{sec:solparticle}

\subsection{Single Particle Motion}

To derive a convenient equation of motion in the presence of a strong magnetic field $\bm B$, we start with the Hamiltonian of a charged particle of species $a$ \citep{Jackson1999},

\be
    H_a(\bm x, \bm p)=\frac{[\bm p - q_a \bm A]^2}{2m_a}+q_a \phi,
    \label{eq:hamiltonian}
\ee

\noindent and its associated Lagrangian,

\begin{equation}
    L_a(\bm x, \bm v)=\left[q_a \bm A(\bm x) + m_a \bm v\right]\cdot \dot{\bm x}-\left(\frac{m_a v^2}{2}+q_a \phi(\bm x)\right),
    \label{eq:lagrangian}
\end{equation}

\noindent where $\bm p = m _a\bm v + q_a \bm A$, $\bm v$ is the particle velocity, $\bm A$ is the magnetic vector potential, $\phi$ is the electrostatic potential, $m_a$ is the mass of the particle and $q_a$ its charge.

We now perform a coordinate transformation from the phase-space coordinates $\bm z = (\bm x, \bm v)$ to the guiding-center coordinates $\bm Z = (\bm R, v_\parallel, \mu, \theta)$ by writing the particle velocity as (see, e.g., \citet{Littlejohn1983a})

\begin{align}
    \bm v &= \bm U+ v_\perp' \bm c,
    \label{eq:GCcoordinates}
\end{align}

\noindent with $\bm U = \bm v_E(\bm R) + v_\parallel \bm b(\bm R)$, $v_\parallel = \bm v \cdot \bm b$, $\bm b = \bm B/B$,  and $\bm v_E = \bm E \times \bm B/B^2$. The gyroangle $\theta = \tan^{-1} \left[\left(\bm v - \bm U \right)\cdot \bm e_2 / \left(\bm v - \bm U \right)\cdot \bm e_1 \right]$ is introduced by defining the right-handed coordinate set $(\bm e_1, \bm e_2, \bm b)$, such that $\bm c = -\bm a \times \bm b=\bm a'(\theta)$, with $\bm a = \cos \theta \bm e_1 + \sin \theta \bm e_2$.
The decomposition in \cref{eq:GCcoordinates} allows us to isolate the high-frequency gyromotion, contained in the $v_\perp' \bm c$ term, from the dominant guiding-center velocity $\bm U$.
The adiabatic moment $\mu$ is defined as

\begin{equation}
    \mu = \frac{m_a v_\perp^{'2}}{2B},
\end{equation}

\noindent whereas the guiding-center position is

\begin{equation}
    \bm R = \bm x - \rho_a \bm a,
    \label{eq:GCx}
\end{equation}

\noindent with $\rho_a = \sqrt{2 m_a  \mu/(q_a^2 B)}$ the Larmor radius. Incidentally, for the case of weakly varying magnetic fields, \cref{eq:GCx} describes the circular motion of a particle around its guiding-center $\bm R$ with radius $\rho_a$, i.e., $(\bm x - \bm R)^2 = \rho_a^2$.

As our goal is to develop a model that describes turbulent fluctuations occurring on a spatial scale longer than the sound Larmor radius $\rho_s$, and a time scale larger than the gyromotion one, we keep terms in the Lagrangian up to $O(\epsilon)$ and order $T_i \lesssim T_e$, which implies
\begin{equation}
    k_{\perp} \rho_i \lesssim \epsilon.
    \label{eq:tiordering}
\end{equation}

We therefore expand the electromagnetic fields around $\bm R$, to first order in $\epsilon$, i.e.,

\begin{equation}
    \phi(\bm x) \simeq \phi(\bm R) + \rho_a \bm a \cdot \nabla_{\bm R} \phi(\bm R),
    \label{eq:FLRexp}
\end{equation}

\noindent and similarly for $\bm A$. In the following, if not specified, the electromagnetic fields and potentials are evaluated at the guiding-center position $\bm R$, and we denote $\nabla=\nabla_{\bm R}$.
In addition, to take advantage of the difference between the turbulent and gyromotion time scales, we use the gyroaveraged Lagrangian $\lb L_a \rb$ to evaluate the plasma particle motion, where the gyroaveraging operator $\lb \chi \rb$ acting on a quantity $\chi(\theta)$ is defined as

\begin{equation}
    \lb \chi \rb = \frac{1}{2\pi}\int_0^{2\pi} \chi (\theta) d\theta,
\end{equation}

\noindent {which is performed at fixed position $\bm x$, as opposed to the gyrokinetic equation that can be obtained by gyroaveraging with $\bm R$ fixed \citep{Hazeltine1968}.}

To evaluate $\lb L_a \rb$ we note that, with the expansion for $\phi$ and $\bm A$, the Lagrangian in \cref{eq:lagrangian} can be expressed as $L_a=L_{0a}+L_{1a}+\tilde L_a$ where $L_{0a}$ is gyroangle independent,

\begin{equation}
    \begin{split}
        L_{0a} &= \left(q_a \bm A + m_a \bm U\right)\cdot \dot{\bm R}-\left(\frac{m_a v_\parallel^2}{2}+\frac{m_a v_E^2}{2}+\mu B+q_a \phi\right),
    \end{split}
    \label{eq:L0}
\end{equation}

\noindent $L_{1a}$ is proportional to $\rho_a^2$ (and hence to $\mu$),

\begin{equation}
    \begin{split}
        &L_{1a} = \rho_a^2 q_a \dot \theta \left(\bm a \cdot \nabla\right) \left(\bm A \cdot \bm c\right)+m_a \rho_a^2 \Omega \dot \theta+{\rho_a \dot \rho_a}\left[q_a \left(\bm a \cdot \nabla\right)\left(\bm A \cdot \bm a\right)\right],
    \end{split}
    \label{eq:L2}
\end{equation}

\noindent and the $\tilde L_{a}$ contribution contains the terms {linearly proportional to $\cos \theta$ or $\sin \theta$} \citep{Cary2009} {which are not present in $\lb L_a \rb$ as $\lb \tilde L_a \rb = 0$}.

We note that $\lb L_{1a} \rb$ can be simplified since $\lb \left(\bm a \cdot \nabla\right) \bm A \cdot \bm c \rb = -\bm b \cdot (\nabla \times \bm A) /2$, and $\lb \left(\bm a \cdot \nabla\right) \bm A \cdot \bm a \rb = \nabla_\perp\cdot\bm{A}/2$. Subtracting the total derivative $-q_a d/dt(\rho_a^2 \nabla_\perp \bm A)/4$ from $\lb L_a \rb$, which does not alter the resulting equations of motion, we redefine the gyroaveraged Lagrangian as

\begin{equation}
    \begin{split}
        \lb L_a \rb &= \left(q_a\bm A + m_a{\bm U}\right) \cdot \dot{\bm R}  - \l(\frac{m_a v_\parallel^2}{2}+\frac{m_a v_E^2}{2} + q_a \phi\r)\\
        &-\mu B\left(1- \frac{\dot \theta}{\Omega_a}\right)
        -\frac{\rho_a^2}{4}\frac{d }{dt}\left[\nabla_\perp\cdot\left(q_a \bm A\right)\right].
    \end{split}
    \label{eq:gyroLag}
\end{equation}

We now order the terms appearing in $\lb L_a \rb$.
As imposed by the Bohm sheath conditions \citep{Stangeby2000}, both electrons and ions stream along the field lines with parallel velocities comparable to the sound speed $c_s = \sqrt{T_e/m_i}$ in the SOL.
The Bohm boundary conditions at the sheath also set the electrostatic potential $e \phi \sim \Lambda T_e$ across the SOL, where $\Lambda = \log \sqrt{m_i/(m_e 2\pi)}\simeq 3$.
Therefore, we keep the $m_a v_E^2/2$ term in the Lagrangian in \cref{eq:gyroLag}, as to take into account the presence of the numerically large factor $\Lambda^2$ in $v_E^2 \sim \epsilon^2 \Lambda^2 c_s^2$.

By neglecting the higher-order terms in \cref{eq:gyroLag}, i.e., $-(\rho_a^2/4){d }\left[\nabla_\perp\cdot\left(q_a \bm A\right)\right]/{dt}$, the expression for the gyroaveraged Lagrangian describing SOL single particle dynamics, up to $O(\epsilon)$, can be written as

\begin{equation}
    \lb L_a \rb = q_a \bm A^* \cdot \dot{\bm R} - q_a \phi^* -\frac{m_a v_\parallel^2}{2}+ \mu \frac{m_a \dot \theta}{q_a}.
    \label{eq:lagSOL}
\end{equation}

\noindent where $q_a\phi^* = q_a\phi+ m_a v_E^2/2+\mu B$, and $q_a\bm A^* = q_a \bm A  + m_a v_\parallel \bm b + m_a \bm v_E$.
The Euler-Lagrange equations applied to the Lagrangian in \cref{eq:lagSOL} for the coordinates $\theta$, $v_\parallel$, and $\mu$, yield, respectively, $\dot \mu = 0$, $v_\parallel = \bm b \cdot \dot{\bm R}$, and $\dot \theta = \Omega_a$. For the ${\bm R}$ coordinate, we obtain

\begin{equation}
    m_a \dot v_\parallel \bm b = q_a (\bm E^* + \dot{\bm R}\times \bm B^*),
    \label{eq:motLag}
\end{equation}

\noindent where the relation $(\nabla \bm A - (\nabla \bm A)^T) \cdot \dot{\bm R} = \dot{\bm R}\times (\nabla \times \bm A)$ has been used, and we defined $\bm E^* = -\nabla \phi^* - \partial_t \bm A^*$, and $\bm B^* = \nabla \times \bm A^*$, with the parallel component of $\bm B^{*}$ given by

\begin{equation}
    B_{\parallel}^* = \bm B^* \cdot \bm b = B + \frac{m_a}{q_a}\bm b \cdot \nabla \times \left(v_\parallel \bm b + \bm v_E\right).
\end{equation}

By projecting \cref{eq:motLag} along $\bm B^*$, we derive $m \dot v_\parallel B_\parallel^*= e \bm E^* \cdot \bm B^*$, while crossing with $\bm b$ yields the guiding-center velocity $\dot{\bm R} B_\parallel^* = v_\parallel \bm B^* + \bm E^* \times \bm B/B$. Using the expressions for the fields $\bm E^{*}$ and $\bm B^{*}$, we obtain 

\be
    \dot{\bm R}  = \bm U+\frac{\bm B}{\Omega_a B_\parallel^*}\times\left(\frac{d \bm U}{dt}+\frac{\mu\nabla B}{m_a}\right),
    \label{eq:GC1}
\ee

and

\be
    m_a \dot v_\parallel = q_a E_\parallel - \mu \nabla_\parallel B + m_a \bm v_E \cdot \frac{d \bm b}{dt}-m_a\mathcal{A},
    \label{eq:GC2}
\ee

\noindent In \cref{eq:GC1,eq:GC2}, {in addition to the time derivatives of the phase-space coordinates $\dot{\bm R}, \dot v_{\parallel}$, that only have an explicit time dependence, we define the total derivative $d/dt$ of a field $\phi(\bm R, t)$ that has an explicit time and $\bm R$ dependence as $d\phi/dt \equiv \partial_t \phi + \bm U \cdot \nabla \phi$}. The $\mathcal{A}$ term represents the higher-order nonlinear terms in $\dot v_\parallel$ that ensure phase-space conservation properties \citep{Cary2009}, and it is given by

\begin{equation}
    \mathcal{A}=\frac{B}{B_\parallel^*}\left(\left.\frac{d \bm U}{dt}\right|_\perp + \mu \nabla_\perp B\right)\cdot \frac{\nabla \times \bm U}{\Omega_a},
\end{equation}

The guiding-center equations of motion (\ref{eq:GC1}) and (\ref{eq:GC2}) satisfy the energy, $E_{gc} = q_a \phi^*+m_a v_\parallel^2/2$ \citep{Cary2009}, and momentum, $\bm P_{gc} = e \bm A^*$ \citep{Cary2009}, conservation laws, i.e.,

\begin{equation}
    \frac{d E_{gc}}{dt} = q_a \frac{\partial \phi^*}{\partial t}- q_a \frac{\partial \bm A^*}{\partial t}\cdot \dot{\bm R},
    \label{eq:enconservation}
\end{equation}

\noindent and

\begin{equation}
    \frac{\partial \bm P_{gc}}{\partial t} = -q_a \nabla \phi^* + q_a \nabla \bm A^* \cdot \dot{\bm R}.
    \label{eq:moconservation}
\end{equation}

In addition, we note that using \cref{eq:GC1,eq:GC2} and Maxwell's equations, a conservation equation for $B_{\parallel}^*$ can be derived:

\begin{equation}
    \frac{\partial B_{\parallel}^*}{\partial t} + \nabla \cdot (\dot{\bm R} B_{\parallel}^*) + \frac{\partial}{\partial v_\parallel}\left(\dot v_\parallel B_{\parallel}^* \right)=0.
    \label{eq:liouvilleGC}
\end{equation}

\subsection{The Guiding-Center Boltzmann Equation}
\label{subsec:gcboltzmann}
The Boltzmann equation for the evolution of the distribution function $f_a(\bm x, \bm v)$ of the particles in $(\bm x, \bm v)$ coordinates is

\be
     \frac{\partial f_a}{\partial t}+\dot{\bm x}\cdot \nabla_{\bm x} f_a + \dot{\bm v}\cdot \nabla_{\bm v} f_a = C(f_a),
     \label{eq:boltzmann}
\ee

\noindent where $C(f_a)=\sum_b C(f_a,f_b) = \sum_b C_{ab}$ is the collision operator.
Because $f_a$ can significantly deviate from a Maxwellian distribution function in the SOL \citep{Battaglia2014}, we consider the {bilinear} Coulomb operator $C_{ab}$ \citep{Balescu1988}, to model collisions between particles of species $a$ and $b$

\be
    \begin{split}
        C_{ab}&=L_{ab} \frac{\partial}{\partial v_i}\l[\frac{\partial^2 G_b}{\partial v_i \partial v_j}\frac{\partial f_a}{\partial v_j}-\frac{m_a}{m_b}\frac{\partial H_b}{\partial v_i}f_a\r],
    \end{split}
    \label{eq:coulombop}
\ee

\noindent with

\be
    \begin{split}
        H_b&=2\int \frac{f_b(\bm v')}{|\bm v - \bm v'|}d\bm v',
    \end{split}
\ee

\noindent and

\be
    \begin{split}
        G_b=\int f_b(\bm v')|\bm v - \bm v'|d\bm v',
    \end{split}
\ee

\noindent the Rosenbluth potentials satisfying $\nabla^2_v G_b = H_b$. In \cref{eq:coulombop} we introduced $L_{ab}=q_a^2 q_b^2 \lambda/(4 \pi \epsilon_0^2 m_a^2)=\nu_{ab} v_{tha}^3/n_b$, where $\lambda$ is the Coulomb logarithm, $\nu_{ab}$ the collision frequency between species $a$ and $b$, and $v_{tha}^2=2 T_a/m_a$.

Taking advantage of the small electron to ion mass ratio, the collision operator between unlike-species can be simplified (see, e.g. \citet{Balescu1988,Helander2002}). 
The electron-ion collisions are modeled by using the operator $C_{ei}(f_e)=C_{ei}^0+C_{ei}^1$, where $C_{ei}^0$ is the Lorentz pitch-angle scattering operator
\be
    \begin{split}
        C_{ei}^0&=\frac{n_i L_{ei}}{8 \pi}\frac{\partial}{\partial \bm c_e}\cdot\l[\frac{1}{c_e}\frac{\partial f_e}{\partial \bm c_e}-\frac{\bm c_e}{c_e^3}\l(\bm c_e \cdot \frac{\partial f_e}{\partial \bm c_e}\r)\r],
    \end{split}
    \label{eq:cei0}
\ee

\noindent and $C_{ei}^1$ the momentum-conserving term

\be
    \begin{split}
        C_{ei}^1&={\frac{\nu_{ei}}{8 \pi v_{the} c_e^3}f_{Me}{\bm u_{i}} \cdot \bm c_e}.
    \end{split}
    \label{eq:cei1}
\ee

\noindent with $\bm c_e = (\bm v - \bm u_e)/v_{the}$.

Ion-electron collisions are modelled with the operator

\begin{equation}
    \begin{split}
        C_{ie}&=\frac{ \bm R_{ei}}{m_i n_i v_{thi}}\cdot \frac{\partial f_i}{\partial \bm c_i}%,\\
        %C_{ie}^1&=
        +\nu_{ei}\frac{n_e}{n_i}\frac{m_e}{m_i}\frac{\partial}{\partial \bm c_i}\cdot\left(\bm c_i f_i%,\\
        %C_{ie}^2&=\nu_{ei}\frac{m_e}{m_i}%
        +\frac{T_e}{T_i}
        %\frac{T_e}{m_i}
        \frac{\partial f_i}{\partial \bm c_i} 
        \right),
    \end{split}
    \label{eq:cie}
\end{equation}

\noindent where $\bm R_{ei}=\int m_e \bm v C_{ei} d\bm v$ is the electron-ion friction force.

We take advantage of \cref{eq:orderingnu} to order the electron collision frequency $\nu_e$ and the ion collision frequency $\nu_i$ as
\begin{equation}
    \frac{\nu_i}{\Omega_i} \sim \sqrt{\frac{m_e}{m_i}}\left(\frac{T_e}{T_i}\right)^{3/2}\epsilon_{\nu} \lesssim \epsilon^2,
    \label{eq:orderingnu2}
\end{equation}

\noindent where we used the relation $\nu_i \sim  \sqrt{{m_e}/{m_i}}(T_e/T_i)^{3/2} \nu_e$. The orderings in \cref{eq:orderingnu2,eq:FLRexp} yield the lower bound in \cref{eq:titebound} for the ion to electron temperature ratio.
We now express the particle distribution function $f_a$ in terms of the guiding-center coordinates {by defining $F_a$, a function of guiding-center coordinates, as}

\begin{equation}
    F_a(\bm R, v_\parallel, \mu, \theta) = f_a(\bm x(\bm R, v_\parallel, \mu, \theta), \bm v(\bm R, v_\parallel, \mu, \theta)).
    \label{eq:fguidF}
\end{equation}

Using the chain rule to rewrite \cref{eq:boltzmann} in guiding-center coordinates, we obtain

\be
     \frac{\partial F_a}{\partial t}+\dot{\bm R}\cdot \nabla F_a + \dot{v_\parallel}\frac{\partial F_a}{\partial v_\parallel} + \dot \mu \frac{\partial F_a}{\partial \mu}+ \dot \theta \frac{\partial F_a}{\partial \theta} = C(F_a),
     \label{eq:boltzmannSS}
\ee

\noindent where $\dot{\bm R}$ and $\dot v_\parallel$ are given by \cref{eq:GC1} and \cref{eq:GC2} respectively, $\dot \theta = \Omega_a$, and $\dot \mu = 0$.
Equation (\ref{eq:boltzmannSS}) can be simplified by applying the gyroaveraging operator. This results in the drift-kinetic equation

\begin{equation}
    \frac{\partial \lb F_a \rb}{\partial t}+ \dot{\bm R} \cdot \nabla\lb F_a \rb + \dot v_{\parallel}\frac{\partial \lb F_a \rb}{\partial v_\parallel} = \lb C(F_a)\rb.
    \label{eq:boltzmannGC1}
\end{equation}

We now write \cref{eq:boltzmannGC1} in a form useful to take gyrofluid moments of the form $\int \lb F_a \rb B dv_\parallel d\mu d\theta$ (see \cref{sec:momenthierarchy}). Using the conservation law in \cref{eq:liouvilleGC} for $B_{\parallel}^{*}$, we can write the guiding-center Boltzmann equation in conservative form as

\begin{equation}
    \begin{split}
        &\frac{\partial (B_{\parallel}^*\lb F_a \rb)}{\partial t}+ \nabla \cdot ( \dot{\bm R} B_{\parallel}^*\lb F_a \rb) + \frac{\partial( \dot v_{\parallel a} B_{\parallel}^*\lb F_a \rb)}{\partial v_\parallel}%\\
        %-\lb F_a \rb \dot{\bm R_a}\cdot {\nabla B}{} = B\lb C(F_a)\rb.
        %&-  B \lb F_a \rb \left\{\frac{\nabla \cdot [B(\dot{\bm R_a}-\bm U)]}{B}-\frac{E_\parallel}{B}[\bm b \cdot (\nabla \times \bm b)]\right\}
        = B_{\parallel}^*\lb C(F_a)\rb. 
    \end{split}
    \label{eq:boltzmannGC}
\end{equation}

Moreover, in order to relate the gyrofluid moments $\int \lb F_a \rb B dv_\parallel d\mu d\theta$ with the usual fluid moments $\int f_a d^3 v$, we estimate the order of magnitude of the gyrophase dependent part of the distribution function $\tilde F_a = F_a - \lb F_a \rb$ where $\lb F_a \rb$ obeys \cref{eq:boltzmannGC1}. The equation for the evolution of $\tilde F_a$ is obtained by subtracting \cref{eq:boltzmannGC1} from the Boltzmann equation, \cref{eq:boltzmannSS}, that is

\begin{equation}
    \frac{\partial \tilde F_a}{\partial t}+\dot{\bm R}\cdot \nabla \tilde F_a + \dot{v_\parallel}\frac{\partial \tilde F_a}{\partial v_\parallel} + \Omega_a \frac{\partial \tilde F_a}{\partial \theta} = C(F_a)-\lb C(F_a)\rb.
    \label{eq:boltztilde}
\end{equation}

Using the orderings in \cref{eq:orderingnu,eq:orderingnu2}, as well as $\partial_t \sim \dot{\bm R} \cdot \nabla \sim \dot v_\parallel \partial_{v_\parallel} \sim \epsilon \Omega_i$ and $\Omega_a \partial_\theta \sim \Omega_a$, the comparison of the leading-order term on the left-hand side of \cref{eq:boltztilde} with the right-hand side of the same equation imply the following ordering for $\tilde F_e$
\begin{equation}
    \frac{\tilde F_e}{\lb F_e \rb} \sim \frac{m_e}{m_i}\epsilon_\nu\lesssim \epsilon^2,
    \label{eq:orderingftildee}
\end{equation}
and $\tilde F_i$
\begin{equation}
    \frac{\tilde F_i}{\lb F_i \rb} \sim \sqrt{\frac{m_e}{m_i}}\left(\frac{T_e}{T_i}\right)^{3/2}\epsilon_\nu \lesssim \epsilon^2.
    \label{eq:orderingftildei}
\end{equation}

To evaluate the leading-order term of $\tilde F_a$, we expand the collision operator $C(F_a) = C_0(\lb F_a\rb) + \epsilon C_1(F_a) + ...$, such that

\begin{equation}
    \tilde F_a \simeq \frac{1}{\Omega_a}\int_0^\theta\left[C_0(\lb F_a \rb )-\lb C_0( \lb F_a \rb)\rb\right]d\theta' + O( \epsilon^3 \lb F_a \rb),
    \label{eq:tildefapp}
\end{equation}

\noindent The relation in \cref{eq:tildefapp} can be further simplified by expanding the $\theta$ dependence of $F_a$ in Fourier harmonics, 

\begin{equation}
    F_a=\sum_m e^{i m \theta} F_{m a},
    \label{eq:fourftilde}
\end{equation}

\noindent so that for $m=0$ we have $\lb F_a \rb = F_{0a}$, and similarly for $C_0(\lb F_a \rb)$

\begin{equation}
    C_0(\lb F_a \rb) = \sum_{m'} e^{i m' \theta} C_{m' a}.
\end{equation}

\noindent We can then write \cref{eq:tildefapp} as

\begin{equation}
    \tilde F_{m a} = \frac{C_{m a}}{i m \Omega_a},
    \label{eq:fmacmafourier}
\end{equation}

\noindent for $m \not=0$.

\section{Moment Expansion}
\label{sec:momentexpansion}

We now derive a polynomial expansion for the distribution function $\lb F_a \rb$ that simplifies the solution of \cref{eq:boltzmannGC}, with the collision operators in Eqs. (\ref{eq:coulombop}) - (\ref{eq:cie}).
This section is organized as follows.
In \cref{section:gcmoment} the Hermite-Laguerre basis is introduced, relating the corresponding expansion coefficients for $\lb F_a \rb$ with its usual gyrofluid moments.
In \cref{section:jifluidexpansion}, we briefly review the fluid moment expansion of the Coulomb collision operator presented in \citet{Ji2006, Ji2008}.
In \cref{section:cabmomentexpansion}, leveraging the work in  \citet{Ji2006, Ji2008}, we expand $C_{ab}$ in terms of the product of the gyrofluid moments, for both like- and unlike-species collisions.
This ultimately gives us the possibility of solving \cref{eq:boltzmannGC} in terms of gyrofluid moments.

\subsection{Guiding-Center Moment Expansion of $\lb F_a \rb$}
\label{section:gcmoment}

To take advantage of the anisotropy introduced by a strong magnetic field, and efficiently treat the left-hand side of \cref{eq:boltzmannGC} where the parallel and perpendicular directions appear decoupled, we express $\lb F_a \rb$ by using {a Hermite polynomial basis expansion for the parallel velocity coordinate \citep{Grad1949,Armstrong1967a,Grant1967,Ng1999,Zocco2011,Loureiro2013a,Parker2015,Schekochihin2016,Tassi2016} and a Laguerre polynomial basis for the perpendicular velocity coordinate {\citep{Zocco2015,Omotani2015,Mandell2017}}. More precisely, we use the following expansion

\be
    \begin{split}
        \lb F_a \rb &=\sum_{p,j=0}^{\infty} \frac{N_a^{pj}}{\sqrt{2^p p!}}F_{Ma}  H_p(s_{\parallel a})L_j(s_{\perp a}^2),
    \end{split}
    \label{eq:gyrof}
\ee

\noindent where the {\textit{physicists'}} Hermite polynomials $H_p$ of order $p$ are defined by \citep{Abramowicz1988}

\begin{equation}
    H_p(x)=(-1)^p e^{x^2}\frac{d^p}{dx^p}e^{-x^2},
\end{equation}

\noindent and normalized via

\begin{equation}
    \int_{-\infty}^{\infty} dx H_p(x) H_{p'}(x) e^{-x^2} = 2^p p! \sqrt{\pi} \delta_{p{p'}},
\end{equation}

\noindent and the Laguerre polynomials  $L_j$ of order $j$ are defined by \citep{Abramowicz1988} 

\begin{equation}
    L_j(x)=\frac{e^x}{j!}\frac{d^j}{dx^j}(e^{-x}x^j),
\end{equation}

\noindent which are orthonormal with respect to the weight $e^{-x}$

\begin{equation}
    \int_{0}^{\infty} dx L_j(x) L_{j'}(x) e^{-x} = \delta_{jj'}.
\end{equation}

Because of the orthogonality of the Hermite-Laguerre basis, the coefficients $N_a^{pj}$ of the expansion in \cref{eq:gyrof} are

\be
    N_a^{pj}=\frac{1}{N_{a}}\int \frac{H_p(s_{\parallel a}) L_j(s_{\perp a}^2) \lb F_a \rb }{\sqrt{2^p p!}}\frac{B}{m_a} d\mu dv_\parallel d\theta,
    \label{eq:gyromoments}
\ee

\noindent and correspond to the guiding-center moments of $\lb F_a \rb$.

In \cref{eq:gyrof}, the shifted bi-Maxwellian is introduced

\be
	F_{Ma}=N_a\frac{e^{-{s_{\parallel a}^2}-s_{\perp a}^2}}{{\pi}^{3/2}v_{th\parallel a} v_{th\perp a}^2},
	\label{eq:bimax}
\ee

\noindent where $s_{\parallel a}$ and $s_{\perp a}$ are the normalized parallel and perpendicular shifted velocities respectively, defined by

\begin{equation}
    s_{\parallel a} = \frac{v_\parallel-u_{\parallel a}}{v_{th\parallel a}},~v_{th\parallel a}^2=\frac{2 T_{\parallel a}}{m_a},
    \label{eq:sparallela}
\end{equation}

\noindent and

\begin{equation}
    s_{\perp a}^2 = \frac{v_\perp^{'2}}{v_{th\perp a}^{2}}=\frac{\mu B}{T_{\perp a}},~v_{th\perp a}^2=\frac{2 T_{\perp a}}{m_a},
    \label{eq:sperpa}
\end{equation}

\noindent {which provide an efficient representation of the distribution function in both the weak ($u_{\parallel a} \ll v_{th a})$ and strong flow ($u_{\parallel a} \sim v_{th a}$) regimes.}

The guiding-center density $N_a$, appearing in \cref{eq:bimax}, the guiding-center fluid velocity $u_{\parallel a}$, in \cref{eq:sparallela}, and the guiding-center parallel $T_{\parallel a}=P_{\parallel a}/N_a$ and perpendicular $T_{\perp a}=P_{\perp a}/N_a$ temperatures in \cref{eq:sparallela,eq:sperpa} are defined as $N_a = ||1||_a$, $N_a u_{\parallel a} = || v_{\parallel}||_a$, $P_{\parallel a} = m_a ||(v_\parallel-u_{\parallel a})^2||_a$, and $P_{\perp a} = ||\mu B ||_a$, where

\begin{equation}
    ||\chi||_a \equiv \int \chi \lb F_a \rb \frac{B}{m_a} d\mu dv_\parallel d\theta.
\end{equation}

The definition of $N_a$, $u_{\parallel a}$, $P_{\parallel a}$, and $P_{\perp a}$ implies that $N_a^{00}=1,~N_a^{10}=0,~N_a^{20}=0,~N_a^{01}=0$ respectively.
Later, we will consider the parallel and perpendicular heat flux, defined as

\begin{align}
        Q_{\parallel a} &= m_a ||(v_\parallel-u_{\parallel a})^3||_a,~Q_{\perp a} = ||(v_\parallel-u_{\parallel a}) \mu B||_a,
        \label{eq:fluidmoments1}
\end{align}

\noindent which are related to the coefficients $N_a^{30}, N_a^{11}$ by %, N_a^{02}, N_a^{21},$ and $N_a^{40}$ by

\be
    \begin{split}
        N_a^{30}&=\frac{Q_{\parallel a}}{\sqrt{3}P_{\parallel a} v_{tha \parallel}},
        ~N_a^{11}=-\frac{\sqrt{2} Q_{\perp a}}{P_{\perp a} v_{tha \parallel}}.
    \end{split}
    \label{eq:kineticmoments1}
\ee

\subsection{Fluid Moment Expansion of the Collision Operator}
\label{section:jifluidexpansion}

A polynomial expansion of the collision operators in \cref{eq:coulombop} was carried out in \citet{Ji2006}, and later extended to effectively take into account finite fluid velocity and unlike-species collisions in \citet{Ji2008}. This allowed expressing $C_{ab}$ as products of fluid moments of $f_a$ and $f_b$. 
We summarize here the main steps of \citet{Ji2006, Ji2008}.

Similarly to \cref{eq:gyrof}, the particle distribution function $f_a$ is expanded as

\be
    f_a = f_{aM} \sum_{l,k=0}^{\infty}\frac{L_k^{l+1/2}(c_a^2) \bm P^{l}(\bm c_a) \cdot {\bm M_a}^{lk}}{\sqrt{\sigma_k^l}},
    \label{eq:faji}
\ee

\noindent where  $f_{aM}=n_a \exp \l(-c_a^2\r)/(\pi^{3/2} v_{tha}^3)$ is a shifted Maxwell-Boltzmann distribution function, and $\bm c_a$ the shifted velocity defined as $\bm c_a=(\bm v - \bm U_a)/v_{tha}$, with $\bm U_a$ the fluid velocity. The fluid variables $n_a, \bm U_a,$ and $T_a$ are defined as the usual moments of the particle distribution function $f_a$, i.e. $n_a = \int f_a d^3 \bm v, { n_a \bm U_a = \int f_a \bm v d^3v, n_a T_a = \int m f_a (\bm v - \bm u_a)^2 d^3v/3}$.

The tensors $\bm P_a^{lk}(\bm c_a)=\bm P^{l}(\bm c_a)L_k^{l+1/2}(c_a^2)$ constitute an orthogonal basis, where $\bm P^l(\bm c_a)$ is the symmetric and traceless tensor

\be
    \begin{split}
    \bm P^l(\bm c_a) &= \sum_{i=0}^{\floor{l/2}}d_i^l S_i^l c_a^{2i}\l\{\bm I^i \bm c_a^{l-2i}\r\},
    \end{split}
\ee

\noindent with $\bm I$ denoting the identity matrix, $\{\bm A^i\}$ denoting the symmetrization of the tensor $\bm A^i$, ${\floor{l/2}}$ denoting the largest integer less than or equal to $l/2$, and the coefficients $d_i^l$ and $S_i^l$ defined by

\begin{equation}
    d_i^l=\frac{(-2)^i(2l-2i)!l!}{(2l)!(l-i)!},
\end{equation}

\noindent and

\begin{equation}
    S_i^l=\frac{l!}{(l-2i)!2^i i!}.
\end{equation}

The tensor $\bm P^l(\bm c_a)$ is normalized via

\begin{equation}
    \int d \bm v \bm P^{n}(\bm v)\bm P^l(\bm v) \cdot \bm M^l g(v) = \bm M^n \delta_{n,l} \sigma_n \int d \bm v v^{2n} g(v),
\label{eq:normplk}
\end{equation}

\noindent with $\sigma_l = l!/[2^l (l+1/2)!]$.
We note that the tensor $\bm A^i$ is formed by $i$ multiplications of the $\bm A$ elements (e.g., if $\bm A$ is a {rank-2 tensor}, $\bm A^3 \equiv \bm A \bm A \bm A$, which in index notation can be written as $(\bm A^3)_{ijlkmn} = A_{ij}A_{lk}A_{mn}$).

In the expansion in \cref{eq:faji}, $L_k^{l+1/2}(x)$ are the associated Laguerre polynomials

\be
    \begin{split}
        L_{k}^{l+1/2}(x)&=\sum_{m=0}^{k}L_{km}^{l} x^m,
    \end{split}
    \label{eq:asslaguerre}
\ee

\noindent normalized via

\begin{equation}
    \int_0^\infty e^{-x} x^{l+1/2} L_k^{l+1/2}(x) L_{k'}^{l+1/2}(x) dx = \lambda_{k}^l\delta_{k,k'}.
\label{eq:normlkl}
\end{equation}

\noindent with $\lambda_{k}^l={(l+k+1/2)!}/{k!}$ and $L_{km}^{l}=[{(-1)^m(l+k+1/2)!}]/[{(k-m)!(l+m+1/2)!m!}]$. The $\sigma_k^l=\sigma_l \lambda_k^l$ term is a normalization factor from the orthogonality relations in \cref{eq:normplk,eq:normlkl}.

Finally, the coefficients of the expansion in \cref{eq:faji} $\bm M_a^{lk}$ are

\be
    \bm M_a^{lk} = \frac{1}{n_a}\int d \bm v f_a \frac{L_k^{l+1/2}(c_a^2) \bm P^{l}(\bm c_a)}{\sqrt{\sigma_k^l}},
    \label{eq:MlkCoulomb}
\ee

\noindent which correspond to the moments of $f_a$  due to the orthogonality relations in \cref{eq:normplk,eq:normlkl}.

By using the expansion in \cref{eq:faji} in the collision operator in \cref{eq:coulombop}, a closed form for $C_{ab}$ in terms of products of $\bm M_a^{lk}$ can be obtained. For like-species collisions it reads

\be
    C_{aa}=\sum_{lkm}\sum_{nqr}\frac{L_{km}^lL_{qr}^n}{\sqrt{\sigma_k^l \sigma_q^n}}c\l(f_a^{lkm},f_a^{nqr}\r),
    \label{eq:JiCab}
\ee

\noindent with

\be
\begin{split}
    c\l(f_a^{lkm},f_a^{nqr}\r)&=f_{aM}\sum_{u=0}^{\text{min}(2,l,n)}\nu_{*aau}^{lm,nr}(c_a^2)\sum_{i=0}^{\text{min}(l,n)-u}d_i^{l-u,n-u}\bm P^{l+n-2(i+u)}(\hat{\bm c_a})\cdot \overline{\bm M_a^{lk}\cdot^{i+u}\bm M_a^{nq}},
\end{split}
    \label{eq:ccjiheld}
\ee

\noindent where $\hat{\bm c_a} = \bm c_a/c_a$, $\cdot^n$ is the $n$-fold inner product (e.g., for the matrix $\bm A = A_{ij}$, $(\bm A \cdot^1 \bm A)_{ij} = \sum_k A_{ki}A_{kj}$), and $\overline{\bm A}$ the traceless symmetrization of $\bm A$ (e.g., $\overline{\bm A} = (A_{ij}+A_{ji})/2-\delta_{ij}\sum_k A_{kk}/3$).
We refer the reader to \citet{Ji2009} for the explicit form of the $\nu_{*abu}^{lm,nr}$ coefficients. The expansion of the unlike-species collisions is reported in \citet{Ji2008}.

\subsection{Guiding-Center Moment Expansion of the Collision Operators}
\label{section:cabmomentexpansion}

In order to apply the gyroaveraging operator to the like-species collision operator $C_{aa}$ in \cref{eq:JiCab}, we expand the fluid moments as $\bm M_{a}^{lk}=\bm M_{a0}^{lk} + \epsilon \bm M_{a1}^{lk} + ...$, aiming at representing the collision operator up to $O(\epsilon_\nu \epsilon)$.
An analytical expression for the leading-order $\bm M_{a0}^{lk}$ in terms of guiding-center moments $N_{a}^{pj}$ can be obtained as follows.
By splitting $f_a = \lb f_a \rb + \tilde f_a$ when evaluating the fluid moments $\bm M_a^{lk}$ according to \cref{eq:MlkCoulomb}, we obtain

\begin{equation}
    \bm M_a^{lk} = \frac{1}{n_a}\int d^3 x' d^3 v' \delta(\bm x' - \bm x)\frac{L_k^{l+1/2}(c_a^{'2}) \bm P^{l}(\bm c'_a)}{\sqrt{\sigma_k^l}}\left(\lb f_a \rb + \tilde f_a\right).
    \label{eq:malkexact1}
\end{equation}

\noindent where the Dirac delta function was introduced to convert the velocity integral into an $(\bm x, \bm v)$ integral that encompasses the full phase-space.
Since the volume element in phase space can be written as $d^3 \bm x d^3\bm v = (B_\parallel^*/m)d \bm R dv_\parallel d \mu d\theta$ \citep{Cary2009}, and defining $\bm x' = \bm R + \rho_a \bm a$, we can write the fluid moments in \cref{eq:malkexact1} as

\begin{equation}
\begin{split}
    \bm M_a^{lk} &= \frac{1}{n_a}\int d \bm R  d v_\parallel d\mu d\theta\frac{B_\parallel^*}{m_a} \delta(\bm x-\bm R -  \rho_a \bm a)\frac{L_k^{l+1/2}(c_a^{'2}) \bm P^{l}(\bm c'_a)}{\sqrt{\sigma_k^l}}\left( \lb F_a \rb + \tilde F_a\right).    
\end{split}
\label{eq:malkexact}
\end{equation}

\noindent where $\lb f_a \rb$ and $\tilde f_a$ in \cref{eq:malkexact1} are written in terms of guiding-center coordinates using \cref{eq:fguidF}.
Neglecting the higher-order $\bm \rho_a$ and $\tilde F_a$ terms, the leading-order fluid moments $\bm M_{a0}^{lk}$ are given by

\begin{equation}
    \bm M_{a0}^{lk} = \frac{1}{n_a}\int d v_\parallel d\mu d\theta\frac{B_\parallel^*}{m_a} \frac{L_k^{l+1/2}(c_a^{'2}) \bm P^{l}(\bm c'_a)}{\sqrt{\sigma_k^l}}\lb F_a \rb .
    \label{eq:malk0exact}
\end{equation}

The $\theta$ integration can be performed by making use of the gyroaveraging formula of the $\bm P^l$ tensor \citep{Ji2009}

\be
    \lb \bm P^l(\bm c_a) \rb = c_a^{l} P_l\l(\xi_a\r) \bm P^l(\bm b),
    \label{eq:Pgyro}
\ee

\noindent where $\xi_a=\bm c_a \cdot \bm b/c_a$ is the pitch angle velocity coordinate, and $P_l$ is a Legendre polynomial defined by

\begin{equation}
    P_l(x)=\frac{1}{2^ll!}\frac{d^l}{dx^l}\left[(x^2-1)^l\right],
\end{equation}

\noindent and normalized via

\begin{equation}
    \int_{-1}^1 P_l(x)P_{l'}(x)dx=\frac{\delta_{ll'}}{l+1/2},
\end{equation}

\noindent yielding

\begin{equation}
    \bm M_{a0}^{lk} = \frac{\bm P^l(\bm b)}{n_a}\int d v_\parallel d\mu  d\theta \frac{B_\parallel^*}{m_a} \frac{L_k^{l+1/2}(c_a^{'2}) c_a^{l} P_l\l(\xi_a\r) }{\sqrt{\sigma_k^l}}\lb F_a \rb .
    \label{eq:malk0exact1}
\end{equation}

Finally, we use the basis transformation

\be
    \begin{split}
        c_a^l P_l(\xi_a)L_k^{l+1/2}(c_a^2)=&\sum_{p=0}^{l+2k}\sum_{j=0}^{k+\floor{l/2}}T_{alk}^{pj}
        H_p(s_{\parallel a})L_j(s_{\perp a}^2),
    \end{split}
    \label{eq:tlkpj}
\ee

\noindent with the inverse

\be
    \begin{split}
        H_p(s_{\parallel a})L_j(s_{\perp a}^2) =& \sum_{l=0}^{p+2j}\sum_{k=0}^{j+\floor{p/2}}\l(T_a^{-1}\r)_{pj}^{lk}
        c_a^l P_l(\xi_a)L_k^{l+1/2}(c_a^2),
    \end{split}
    \label{eq:tminus1pjlk}
\ee

\noindent to obtain an expression in terms of the Hermite-Laguerre basis.
A numerical evaluation of  $T_{alk}^{pj}$ and $\l(T_a^{-1}\r)_{pj}^{lk}$ was carried out in \citet{Omotani2015}. Instead, in Appendix \ref{app:tlkpj}, we derive {the} analytic expressions {of both $T_{alk}^{pj}$ and $\l(T_a^{-1}\r)_{pj}^{lk}$}.

Using the definition of guiding-center moments $N_a^{pj}$ in \cref{eq:gyromoments}, the leading-order fluid moment $\bm M_{a0}^{lk}$ is then given by

\be
    n_a \bm M_{a0}^{lk} = N_a \bm P^l(\bm b) \mathcal{N}_a^{lk},
    \label{eq:CoulDKmom}
\ee

\noindent where we define
\begin{equation}
    \mathcal{N}_a^{lk} = \sum_{p=0}^{l+2k}\sum_{j=0}^{k+\floor{l/2}}T_{alk}^{pj}{N}_a^{pj}\sqrt{\frac{2^p p!}{\sigma_k^l}}.
\end{equation}

The leading-order part $C_{aa0}$ of the collision operator $C_{aa}$ can be calculated by approximating $M_{a}^{lk}$ appearing in \cref{eq:ccjiheld} with $\bm M_{a0}^{lk}$.
For the ions, the largest contribution to $\bm M_{i}^{lk}-\bm M_{i0}^{lk}$ is of order $\epsilon$ and it is given by the $\rho_i$ appearing in \cref{eq:malkexact} (the $\tilde F_i$ correction is smaller since $\tilde F_i \lesssim \epsilon^2 \lb F_i \rb$, see \cref{eq:orderingftildei}). Therefore, by using the ordering in \cref{eq:orderingnu2}, the largest correction to $C_{ii0}$ is $O(\sqrt{m_e/m_i} \epsilon \epsilon_\nu)$. The correction to $C_{ee0}$ is of the same order. It follows that we can approximate $C_{aa}$ appearing in \cref{eq:ccjiheld} with $C_{aa0}$ to represent the collision operator up to $O(\epsilon_{\nu} \epsilon)$.

As an aside, we note that the relationship between the guiding-center and fluid moments in \cref{eq:CoulDKmom} provides, for the indices $(l,k)=(0,0)$,

\begin{align}
    n_a &= N_a,
\end{align}

\noindent while, for $(l,k)=(0,1)$,

\begin{align}
    T_a = \frac{T_{\parallel a}+2 T_{\perp a}}{3}.
\end{align}

Moreover, the $(l,k)=(2,0)$ moment provides a relationship useful to express the viscosity tensor $\bm \Pi_a = \int (\bm c_a \bm c_a - c_a^2 \bm I) f_a d\bm v$

\begin{align}
    \bm \Pi_a = \bm b \bm b N (T_{\parallel a} - T_{\perp a}),
\label{eq:stressdk}
\end{align}

\noindent and $(l,k)=(1,1)$ gives

\begin{align}
    \bm q_a &= \left(\frac{Q_{\parallel a}}{2}+Q_{\perp a}\right)\bm b,
\label{eq:heatfluxdk}
\end{align}

\noindent with $\bm q_a$ the heat flux density $\bm q_a = m \int \bm c_a c_a^2 f_a d \bm v/2$.

In order to express the Boltzmann equation, \cref{eq:boltzmannGC}, in terms of the guiding-center moments $N_a^{pj}$, we evaluate the guiding-center moments of $\lb C_{aa} \rb$, namely
\be
    \begin{split}
         C_{aa}^{pj}=\frac{1}{N_a} \int \lb C_{aa0} \rb  \frac{H_p(s_{\parallel a}) L_j(s_{\perp a}^2)}{\sqrt{2^p p!}} \frac{B}{m_a} dv_{\parallel} d\mu d\theta.
        \label{eq:CoulIntGyro}
    \end{split}
\ee

By using the gyroaveraging property (\ref{eq:Pgyro}) of $\bm P^{l}(\bm c_a)$ in the like-species operator in \cref{eq:JiCab,eq:ccjiheld} (with $\bm M_{a}^{lk} = \bm M_{a0}^{lk}$), and the relation (\ref{eq:CoulDKmom}) between $\bm M_{a0}^{lk}$ and $N_{a}^{pj}$, the gyroaveraged collision operator coefficients $\lb c\l(f_a^{lkm},f_a^{nqr}\r) \rb$ are given by

\be
    \begin{split}
    &\lb c\l(f_a^{lkm},f_a^{nqr}\r)\rb=f_{aM}\sum_{u=0}^{\text{min}(2,l,n)}\nu_{*aau}^{lm,nr}(c_a^2)\sum_{i=0}^{\text{min}(l,n)-u}d_i^{l-u,n-u} P_{l+n-2(i+u)}(\xi) \mathcal{N}_a^{lk}\mathcal{N}_a^{nq}\mathcal{P}^{l,n}_{i+u},
    \end{split}
    \label{eq:JiCabGyro}
\ee

\noindent with $\mathcal{P}^{l,n}_{i+u}=\bm{P}^{l+n-2(i+u)}\cdot \overline{\bm P^{l}\cdot^{i+u}{\bm P}^{n}}$.

Using the basis transformation of \cref{eq:tminus1pjlk} to express ${H_p(s_{\parallel a}) L_j(s_{\perp a}^2)}$ in \cref{eq:CoulIntGyro} in terms of $c_a^l P_l(\xi_a)L_k^{l+1/2}(c_a^2)$, and performing the resulting integral, we obtain

\be
    \begin{split}
        C_{aa}^{pj}=&\sum_{l,k}\sum_{n,q}\sum_{u=0}^{\text{min}(2,l,n)}\sum_{i=0}^{\text{min}(l,n)-u}\sum_{e=0}^{p+2j}\sum_{f=0}^{j+\floor{p/2}}\sum_{g=0}^f\sum_{m=0}^k\sum_{r=0}^q\\
        &\frac{L_{km}^lL_{qr}^n L_{fg}^e d_i^{l-u,n-u}}{\sqrt{\sigma_k^l \sigma_q^n} (e+1/2)4 \pi}\frac{C_{*aau}^{eg,lm,nr}}{\sqrt{2^p p!}}\delta_{e,l+n-2(i+u)}{\l(T^{-1}\r)}_{pj}^{ef} \mathcal{N}_a^{lk}\mathcal{N}_a^{nq}\mathcal{P}^{l,n}_{i+u},
    \end{split}
    \label{eq:caapjexact}
\ee

\noindent with $C_{*aabu}^{jw,lm,nr}=\int d\bm v c_a^{2w+j}f_{Ma}\nu_{*aau}^{lm,nr}$.

We now turn to the electron-ion collision operator, $C_{ei} = C_{ei}^0 + C_{ei}^1$, with $C_{ei}^0$ given by \cref{eq:cei0} and $C_{ei}^1$ given by \cref{eq:cei1}.
As the basis $ L_k^{l+1/2} \bm P^l(\bm c_a)$ is an eigenfunction of the Lorentz pitch-angle scattering operator $C_{ei}^0$ with eigenvalue $-l(l+1)$ \citep{Ji2008}, we write $C_{ei}^0$ as

\be
    C_{ei}^0=-\sum_{l,k}\frac{n_i L_{ei}}{8 \pi c_e^3}\frac{l(l+1)f_{eM}}{\sqrt{\sigma_k^l}}L_k^{l+1/2}(c_e^2) \bm P^l(\bm c_e) \cdot {\bm M_e}^{lk}.
    \label{eq:cei0eig}
\ee

Similarly to like-species collisions, we approximate $\bm M_{e}^{lk} \simeq \bm M_{e0}^{lk}$ in \cref{eq:cei0eig}, representing $C_{ei}^0$ accurately up to $O(\epsilon_\nu \epsilon)$.
Using the basis transformation in \cref{eq:tminus1pjlk} and the gyroaverage property of $\bm P^l (\bm c_a)$ in \cref{eq:Pgyro}, we take guiding-center moments of $C_{ei}$ of the form (\ref{eq:CoulIntGyro}), and obtain

\be
\begin{split}
    C_{ei}^{pj} = -\frac{\nu_{ei}}{8 \pi^{3/2}}&
    \sum_{l=0}^{p+2j}\sum_{f=0}^{j+\floor{p/2}}
    \frac{{\l(T^{-1}_e\r)}_{pj}^{lf}}{\sqrt{2^p p!}}\left[\sum_{k=0}^\infty A_{ei}^{lf,k} \mathcal{N}_e^{lk} -\delta_{l,1}\frac{{u_{\parallel i}}}{v_{the}}\frac{{16}}{3 } \frac{\Gamma(f+3/2)}{f!\sqrt{\pi}}\right],
\end{split}
\label{eq:ceipj}
\ee

\noindent where the $A_{ei}$ coefficients are given by

\be
    \begin{split}
        A_{ei,0}^{lf,k}=&\frac{l(l+1)}{l+1/2}|\bm P^l(\bm b)|^2 \sum_{m=0}^f \sum_{n=0}^k \frac{L_{fm}^l L_{kn}^l}{\sqrt{\sigma_k^l}} {(l+m+n-1)!}.
    \end{split}
\ee

Finally, for the ion-electron collision operator, $C_{ie}$, we neglect $O(\sqrt{m_e/m_i}\epsilon_\nu \epsilon)$ corrections by approximating $F_i \simeq \lb F_i \rb$, and use the transformation in \cref{eq:GCcoordinates} to convert the $C_{ie}$ operator in \cref{eq:cie} to guiding-center variables, yielding

\begin{align}
    C_{ie} &= \frac{\bm R_{ei}}{m_i n_i v_{th i }}\cdot\left[\bm c_\perp \frac{m_i v_{thi}^2}{B}\frac{\partial \lb F_i \rb}{\partial \mu}+\bm b \frac{\partial \lb F_i \rb}{\partial c_{\parallel i}}\right]+\nu_{ei}\frac{m_e}{m_i}\frac{n_e}{n_i}\bigg[3 \lb F_i \rb\nonumber\\
    &\l. + c_{\parallel i} \frac{\partial \lb F_i \rb}{\partial c_{\parallel i}} +2\mu \frac{\partial \lb F_i \rb}{\partial \mu}+ \frac{T_e}{2T_i}\frac{\partial^2 \lb F_i \rb}{\partial c_{\parallel i }^2}+\frac{2 T_e}{B} \frac{\partial}{\partial \mu}\left(\mu \frac{\partial \lb F_i \rb}{\partial \mu}\right)\right].
    \label{eq:cie11}
\end{align}

By evaluating $\bm R_{ei}$ at the guiding-center position $\bm R$, we write $\bm R_{ei} \cdot \bm b = N_e m_e v_{th\parallel e} C_{ei}^{10}/\sqrt{2} + O(\sqrt{m_e/m_i}\epsilon_\nu \epsilon)$ and gyroaverage \cref{eq:cie11}, yielding

\begin{equation}
    \begin{split}
        \lb C_{ie} \rb&=\frac{C_{ei}^{10}}{\sqrt{2}}\frac{m_e}{m_i} \frac{N_e}{n_i} \frac{v_{th\parallel e}}{v_{th\parallel i}} \frac{\partial \lb F_i\rb }{\partial s_{\parallel}}+\nu_{ei}\frac{m_e}{m_i}\frac{n_e}{n_i}\bigg[3 \lb F_i \rb+\\
        %\lb C_{ie}^1 \rb&=
        &\l. s_{\parallel i} \frac{\partial \lb F_i \rb}{\partial s_{\parallel i}}+2\mu \frac{\partial \lb F_i \rb}{\partial \mu}+ \frac{T_e}{2 T_{\parallel i}}\frac{\partial^2 \lb F_i \rb}{\partial s_{\parallel i}^2} +\frac{2 T_e }{B}\frac{\partial}{\partial \mu}\left(\mu \frac{\partial \lb F_i \rb}{\partial \mu}\right) \r],
    \end{split}
    \label{eq:gyrocie}
\end{equation}

\noindent where we used $c_{\parallel i}^2 = s_{\parallel i}^2 T_{\parallel i}/T_{i}$.
Taking guiding-center moments of the form (\ref{eq:CoulIntGyro}) of $\lb C_{ie} \rb$ in \cref{eq:gyrocie}, we obtain

\begin{equation}
    C_{ie}^{pj}=\nu_{ei}\frac{m_e}{m_i}\sum_{lk}B_{lk}^{pj}N_{i}^{lk},
    \label{eq:ciepj}
\end{equation}

\noindent with

\begin{equation}
\begin{split}
    B_{lk}^{pj}&=2j\delta_{lp}\delta_{kj-1}\l(1-\frac{T_e}{T_{\perp i}}\r)-\sqrt{ p}\frac{v_{th\parallel e}}{v_{th\parallel i}}\frac{C_{ei}^{10}}{\nu_{ei}}\delta_{lp-1}\delta_{kj}\\
    &-(p+2j)\delta_{lp}\delta_{kj}+\sqrt{ p (p-1)}\delta_{l p-2}\delta_{kj}\left(\frac{T_e}{T_{\parallel i}}-1\right).
\end{split}
\end{equation}

\section{Moment Hierarchy}
\label{sec:momenthierarchy}

In this section, we derive a set of equations that describe the evolution of  the guiding-center moments $N_a^{pj}$, by integrating in guiding-center velocity space the conservative form of the Boltzmann equation, \cref{eq:boltzmannGC}, with the weights $H_p(s_{\parallel a})L_j(s_{\perp a}^2)$.
First, we highlight the dependence of $\dot{\bm R}$ and $\dot v_{\parallel}$ on $s_{\parallel a}$ and $s_{\perp a}^2$ by rewriting the equations of motion as

\begin{equation}
\begin{split}
    \dot{\bm R} &= \bm U_{0 a} + \bm U_{p a}^* + s_{\perp a}^2 \bm U_{\nabla B a}^* + s_{\parallel a}^2 \bm U_{k a}^* + s_{\parallel a}(v_{th\parallel a} \bm b + \bm U_{p a}^{*th}),
\end{split}
\label{eq:rdotGCform}
\end{equation}

\noindent and

\begin{equation}
\begin{split}
    m_a \dot{v}_\parallel &= F_{\parallel a}-s_{\perp a}^2 F_{M a} +s_{\parallel a} F_{p a}^{th}-m_a \mathcal{A}.
\end{split}
\label{eq:vparGCform}
\end{equation}

In \cref{eq:rdotGCform,eq:vparGCform}, $\bm U_{0 a} = \bm v_E + u_{\parallel a} \bm b$ is the lowest-order guiding-center fluid velocity, $\bm U_{\nabla B a}^* = (T_{\perp a}/m_a)(\bm b \times \nabla B/\Omega_a^{*} B)$ is the fluid grad-B drift, with $\Omega_a^{*} = q_a B_{\parallel}^* / m_a$, $\bm U_{ka}^{*} = (2 T_{\parallel a}/m_a)(\bm b \times \bm k/\Omega_a^*)$ is the fluid curvature drift with $\bm k = \bm b \cdot \nabla \bm b$, $\bm U_{pa}^* = ({\bm b}/{\Omega_a^*})\times d_0 \bm U_{0 a}/dt$ is the fluid polarization drift, $F_{\parallel a} = q_a E_\parallel+m_a\bm v_E \cdot d_0 \bm b/{dt}$, $F_{M a} = {T_{\perp a}}{}\nabla_\parallel \ln B$ is the mirror force, and both $\bm U_{p a}^{*th}$ and $F_{p a}^{th}$ are related to gradients of the electromagnetic fields

\begin{equation}
\begin{split}
    \bm U_{p a}^{*th} &= v_{th\parallel a}\frac{\bm b}{\Omega_a^*}\times \left(\bm b \cdot \nabla \bm v_E+\bm v_E \cdot \nabla \bm b + 2 u_{\parallel a} \bm k\right),\\
    F_{p a}^{th} &=  m_a v_{th\parallel a} \bm b \cdot \left(\frac{\bm k \times \bm E}{B}\right).
\end{split}
\end{equation}

The fluid convective derivative operator is defined as 

\begin{equation}
    \frac{d_{0 a}}{dt} = \partial_t + \bm U_{0 a} \cdot \nabla.  
    \label{eq:convdev0}
\end{equation}

Next, to obtain an equation for the moment $N_a^{pj}$, we apply the guiding-center moment operator

\begin{equation}
\begin{split}
    ||\chi||_a^{*pj} &=\frac{1}{N_a B} ||\chi H_p(s_{\parallel a}) L_j(s_{\perp a}^2) B_{\parallel}^*||\\
    &= \frac{1}{N_a}\int \chi \frac{B_\parallel^*}{m_a} \lb F_a \rb \frac{H_p(s_{\parallel a}) L_j(s_{\perp a}^2)}{\sqrt{2^p p!}}  dv_\parallel d\mu d\theta,
\end{split}
\end{equation}

\noindent to Boltzmann's equation, \cref{eq:boltzmannGC}. By defining $|| 1 ||_a^{*pj} = \overline{N}_a^{pj}$ such that

\begin{equation}
\begin{split}
    \overline{N}_a^{pj} &= N_a^{pj}\left(1+\frac{\bm b \cdot \nabla \times \bm v_E}{\Omega_a}+u_{\parallel a} \frac{\bm b \cdot \nabla \times \bm b}{\Omega_a}\right)\\
    &+ v_{th\parallel a}\frac{\bm b \cdot \nabla \times \bm b}{\sqrt{2}\Omega_a}\left(\sqrt{p+1}N_a^{p+1~j}+\sqrt{p}N_a^{p-1~j}\right),
\end{split}
\label{eq:overlinenapj}
\end{equation}

\noindent and

\begin{equation}
\begin{split}
    \frac{d_a^{*pj}}{dt}=\overline{N}_a^{pj}\frac{\partial}{\partial t}+\left|\left|\dot{\bm R}\right|\right|_a^{*pj} \cdot \nabla,
\end{split}
\end{equation}

\noindent the drift-kinetic moment hierarchy conservation equation for species $a$ is

\be
    \begin{split}
        \frac{\partial \overline{N}_a^{pj}}{\partial t} + \nabla \cdot \l|\l|{\dot{\bm R}}\r|\r|_a^{*pj}-\frac{\sqrt{2 p}}{v_{th\parallel a}} \l|\l|\dot v_\parallel\r|\r|_a^{*p-1j} +\mathcal{F}_a^{pj}= \sum_b C_{ab}^{pj},
    \end{split}
    \label{eq:finalDKE}
\ee

\noindent where we identify the fluid operator

\be
    \begin{split}
        \mathcal{F}_a^{pj} &= \frac{d_a^{*pj}}{dt}\ln\left(N_a T_{\parallel a}^{p/2} T_{\perp a}^jB^{-j}\right)+\frac{\sqrt{2p}}{v_{th\parallel a}}\frac{d^{*p-1 j}u_{\parallel a}}{dt}\\
        &+\frac{\sqrt{p(p-1)}}{2}\frac{d_a^{*p-2 j}}{dt}\ln T_{\parallel a}-j\frac{d_a^{*pj-1}}{dt}\ln\left(\frac{ T_{\perp a}}{B}\right),
    \end{split}
    \label{eq:finalDKEF}
\ee

\noindent since it is the key item that describes the evolution of the guiding-center fluid properties $N_a, u_{\parallel a}, P_{\perp a},$ and $P_{\parallel a}$ (see \cref{sec:fluidmodel}).

The guiding-center moments of the particle's equations of motion are given by

\be
    \begin{split}
        \l|\l|{\dot{\bm R}}\r|\r|_a^{*pj}&= \sum_{l,k}\left(\bm U_{0 a}\delta_{pl}\delta_{jk}  + v_{th\parallel a}\bm b\mathcal{V}_{lk}^{1pj}\right)\overline{N}_a^{lk}\\
        & +\left(\bm U_{pa}\delta_{pl}\delta_{jk} + \bm U_{pa}^{th} \mathcal{V}_{lk}^{1pj}+ \bm U_{\nabla B a}\mathcal{M}_{lk}^{pj} + \bm U_{k a}\mathcal{V}_{lk}^{2pj}\right) N_a^{lk},
    \end{split}
    \label{eq:finalDKE3}
\ee
\be
    \begin{split}
        m_a\l|\l|{\dot v_\parallel}{}\r|\r|_a^{*pj}&=\sum_{l,k}\left[F_{\parallel a} \delta_{p,l}\delta_{j,k} +  F_{p a}^{th} \mathcal{V}_{lk}^{1pj} +F_{M a}\mathcal{M}_{lk}^{pj}\right] \overline{N}_a^{lk}+m_a\l|\l| \mathcal{A}\r|\r|_a^{*pj}.
    \end{split}
    \label{eq:finalDKE4}
\ee

\noindent where the phase-mixing operators read

\begin{align}
    \mathcal{V}_{lk}^{1pj}&=\l(\sqrt{\frac{p+1}{2}}\delta_{p+1, l}+\sqrt{\frac{p}{2}}\delta_{p-1 ,l}\r)\delta_{k,j},\label{eq:finalDKE5}\\
    \mathcal{V}_{lk}^{2pj}&=\l[\delta_{p,l}\l(p+\frac{1}{2}\r)+\frac{\sqrt{(p+2)(p+1)}}{2}{\delta_{p+2 ,l}}{}+ \frac{\sqrt{p(p-1)}}{2}\delta_{p-2,l} \r]\delta_{j,k},\label{eq:finalDKE6}\\
    \mathcal{M}_{lk}^{pj}&=(2j+1)\delta_{p,l}\delta_{j,k}-(j+1)\delta_{p,l}\delta_{j+1, k}-j\delta_{p,l}\delta_{j-1 ,k}.\label{eq:finalDKE7}
\end{align}

\noindent The expressions of $\bm U_{p a}, U_{\nabla B_a}, U_{p a}^{th}$, and $\bm U_{k a}$ are derived from $\bm U_{p a}^{*}, U_{\nabla B a}^{*}, U_{p a}^{*th}$, and $\bm U_{k a}^{*}$ by replacing $\Omega_a^*$ with $\Omega_a$. The expression of $\l|\l| \mathcal{A}\r|\r|^{*pj}$ can be found in Appendix \ref{app:collcoef}.

{Similar moment hierarchy models (with uniform magnetic fields) have been numerically implemented, and successfully compared with their kinetic counterpart \citep{Paskauskas2009,Loureiro2015,Schekochihin2016,Groselj2017}, and even shown to be more efficient than other velocity discretization techniques in the same region of validity \citep{Camporeale2016}. Equation (\ref{eq:finalDKE}) generalizes such models to spatially varying fields and full Coulomb collisions, while retaining phase-mixing operators that couple nearby Hermite and Laguerre moments and providing a close form for the projection of the Coulomb operator in velocity space. We also note that the use of shifted velocity polynomials in the Hermite-Laguerre basis, which gives rise to the fluid operator $\mathcal{F}_a^{pj}$, allows us to have an efficient representation of the distribution function both in the weak ($u_{\parallel a} \ll v_{th a})$ and strong flow ($u_{\parallel a} \sim v_{th a}$) regimes. As we will see in \cref{sec:fluidmodel}, the fluid operator $\mathcal{F}_a^{pj}$ generates the lowest order fluid equations, as it is present even if all kinetic moments $N_a^{pj}$ (except $N_a^{00}$) are set to zero.}

\section{Poisson's Equation}
\label{sec:poisson}

We use Poisson's equation to evaluate the electric field appearing in the moment hierarchy equation, \cref{eq:finalDKE}.
In $(\bm x, \bm v)$ coordinates, Poisson's equation reads

\begin{equation}
\begin{split}
        \epsilon_0 \nabla \cdot \bm E &= \sum_a q_a n_a=\sum_a q_a \int f_a d^3 v.
\end{split}
\label{eq:fmoment}
\end{equation}

Following the same steps used to derive \cref{eq:malkexact} from \cref{eq:MlkCoulomb}, we can write Poisson's equation, \cref{eq:fmoment}, as

\begin{equation}
    \epsilon_0 \nabla \cdot \bm E = \sum_a q_a \int d^3 \bm R dv_\parallel d\mu d\theta \frac{B_\parallel^*}{m} \delta(\bm R + \rho_a \bm a - \bm x)F_a(\bm R, v_\parallel, \mu, \theta).
    \label{eq:poissonexact1}
\end{equation}

\noindent Equation (\ref{eq:poissonexact1}) shows that all particles that have a Larmor orbit crossing a given point $\bm x$, give a contribution to the charge density at this location.

Performing the integral over $\bm R$ and introducing the Fourier transform $F_a(\bm x - \rho_a \bm a,v_\parallel, \mu, \theta) = \int d^3\bm k F_a(\bm k,v_\parallel, \mu, \theta) e^{-i \bm k \cdot \bm x} e^{i \rho_a \bm k \cdot \bm a}$, \cref{eq:poissonexact1} can be rewritten as

\begin{equation}
    \epsilon_0 \nabla \cdot \bm E = \sum_a q_a \int  dv_\parallel d\mu d^3 \bm k d\theta \frac{B_\parallel^*}{m_a} F_a(\bm k, v_\parallel, \mu, \theta)e^{-i \bm k \cdot \bm x} e^{i \rho_a \bm k \cdot \bm a}.
    \label{eq:poissonexact2}
\end{equation}

To perform the $\bm k$ integration, we use the cylindrical coordinate system $(k_{\perp}, \alpha, k_\parallel)$, expressing $\bm k = k_\perp (\cos \alpha \bm e_1+\sin\alpha \bm e_2)+k_\parallel \bm b$, such that $\bm k \cdot \bm a = k_\perp \cos (\theta-\alpha)$. This coordinate system allows us to express $e^{i  \rho_a \bm k \cdot \bm a}$ in \cref{eq:poissonexact2} in terms of Bessel functions. {Indeed}, $e^{i k_\perp \rho_a \cos (\theta-\alpha)} = J_0(k_\perp \rho_a) + 2 \sum_{l=1}^{\infty} J_l(k_\perp \rho_a) i^l \cos [l (\theta-\alpha)]$ \citep{Abramowitz1972}, where $J_l(k_\perp \rho_a)$ is the Bessel function of the first kind of order $l$. We can then write

\begin{equation}
\begin{split}
    \epsilon_0 \nabla \cdot \bm E = \sum_a &q_a \int  dv_\parallel d\mu d\theta \frac{B_\parallel^*}{m_a}  \left(  \Gamma_0[F_a]+  2\sum_{l=1}^{\infty} i^l \Gamma_l[F_a \cos[l(\theta-\alpha)] ]\right).    
\end{split}
\label{eq:poissonotexact3}
\end{equation}

\noindent where the Fourier-Bessel operator $\Gamma_l[f]$ is defined as

\begin{equation}
    \Gamma_l[F_a(\bm k, v_\parallel, \mu, \theta)] \equiv \int d^3 \bm k J_l(k_\perp \rho_a) F_a(\bm k, v_\parallel, \mu, \theta) e^{-i \bm k \cdot \bm x}.
\end{equation}

Introducing the Fourier decomposition of $\tilde F_a$, \cref{eq:fmacmafourier}, in  \cref{eq:poissonotexact3}, we obtain

\begin{equation}
\begin{split}
    \epsilon_0 \nabla \cdot \bm E =& \sum_a q_a \int  dv_\parallel d\mu \frac{B_\parallel^*}{m}  \left[ \Gamma_0[\lb F_a \rb] +  2\pi\sum_{l=1}^{\infty} \frac{i^{l-1}}{l\Omega_a}\Gamma_l[C_{l a}e^{i l \alpha}+C_{-l a}e^{-i l \alpha}]\right],
\end{split}
\label{eq:poissonpresqueexact}
\end{equation}

\noindent where the $\theta$ integration was performed by using the identity $\int_0^{2\pi} e^{i\theta(l-m)} d\theta = 2\pi \delta(l-m)$.
Notice that $\int_0^{2\pi}\Gamma_0[F_a]d\theta/2\pi = \Gamma_0(\lb F_a \rb)$, and corresponds to the $J_0(k_\perp \rho_a)$ operator used in most gyrofluid closures \citep{Hammett1992a,Snyder2001,Madsen2013}, and in the gyrokinetic Poisson equation \citep{Lee1983,Dubin1983a}.

We now order the terms appearing in \cref{eq:poissonpresqueexact}. Using the Taylor series expansion of a Bessel function $J_l(x)$ of order $l$ \citep{Abramowitz1972}, we find
\begin{equation}
    \Gamma_0[\lb F_a \rb] \sim \left[1 - \frac{(k_{\perp} \rho_a)^2}{4} + O(\epsilon^{4})\right]\lb F_a \rb,
\end{equation}
\noindent while using the orderings of $\nu_e$ and $\nu_i$ in \cref{eq:orderingnu,eq:orderingnu2}
\begin{equation}
    \frac{\Gamma_l[C_{la}]}{\Omega_a} \lesssim \epsilon_\nu\epsilon^{l+1} \lb F_a \rb.
\end{equation}

\noindent for $l \ge 1$.

Consistently with \cref{section:cabmomentexpansion}, we neglect the $l\ge1$ collisional  terms, therefore representing Poisson's equation up to $O(\epsilon_\nu \epsilon)$.
For the derivation of an higher-order Poisson equation, the treatment of finite $l\ge1$ collisional effects are presented in Appendix \ref{app:poisson}.
Taylor expanding $J_0(x) \simeq 1-x^2/4$, Poisson's equation reads

\begin{equation}
\begin{split}
    \epsilon_0 \nabla \cdot \bm E=&\sum_a q_a\left[N_{a}\left(1+\frac{\bm b \cdot \nabla \times \bm b}{\Omega_a} u_{\parallel a}+\frac{\bm b \cdot \nabla \times \bm v_E}{\Omega_a}\right) +\frac{1}{2m_a}\nabla_\perp^2 \left(\frac{P_{\perp a}}{\Omega_a^2}\right)\right].
\end{split}
    \label{eq:poissonfin2}
\end{equation}

\section{Collisional Drift-Reduced Fluid Model}
\label{sec:fluidmodel}

The infinite set of equations that describe the evolution of the moments of the distribution function, \cref{eq:finalDKE}, and Poisson's equation, \cref{eq:poissonfin2}, constitute the drift-reduced model, which is valid for distribution functions arbitrarily far from equilibrium. For practical purposes, a closure scheme must be provided in order to reduce the model to a finite number of equations.
In this section, we derive a closure in the high-collisionality regime. For this purpose, we first state in \cref{sec:fluideqs} the evolution equations for the fluid moments (i.e. $n_a, u_{\parallel a}, T_{\parallel a}, T_{\perp a}, Q_{\parallel a}$ and $Q_{\perp a}$), that correspond to the lowest-order indices of the moment hierarchy equation. Then, in \cref{sec:highcoll}, we apply a prescription for the higher-order parallel and perpendicular moment equations that allows a collisional closure for $Q_{\parallel a}$ and $Q_{\perp a}$ in terms of $n_a, u_{\parallel a}, T_{\parallel a}$ and $T_{\perp a}$.
The nonlinear closure prescription used here, {sometimes called \textit{semi-collisional closure} \citep{Zocco2011}}, can be employed at arbitrary collisionalities by including a sufficiently high number of moments {(indeed, it was used in \citet{Loureiro2015} to consider low-collisionality regimes). It also allows us to retain the non-linear collision contributions inherent to a full-F description that may have the same size as its linear contributions, as pointed out in \citet{Catto2004}.}

\subsection{Fluid Equations}
\label{sec:fluideqs}

We first look at the $(p,j)=(0,0)$ case of \cref{eq:finalDKE}. Noting that $\overline{N}_a^{00} = 0$ and $C_{ab}^{00}=0$, we obtain

\begin{equation}
    \nabla \cdot \left|\left|\dot{\bm R}\right|\right|^{*00}_a + \mathcal{F}_a^{00}=0.
    \label{eq:cont1}
\end{equation}

Evaluating $\left|\left|\dot{\bm R}\right|\right|^{*pj}_a$ in \cref{eq:finalDKE3} and $\mathcal{F}_a^{pj}$ in \cref{eq:finalDKEF}, for $(p,j)=(0,0)$, \cref{eq:cont1} yields the continuity equation

\begin{equation}
    \frac{d_a^0 N_a}{dt} + \frac{d_{0 a}}{dt}\left(\frac{N_a\nabla_\perp^2 \phi}{\Omega_a B}\right) = -N_a \nabla \cdot \bm u_{0 a} - \frac{N_a\nabla_\perp^2 \phi}{\Omega_a B}\nabla \cdot \bm U_{0 a}.
    \label{eq:continuity}
\end{equation}

The upper convective derivative ${d_a^0}/{dt}$, defined by

\begin{equation}
    \frac{d_a^0}{dt}=\frac{\partial}{\partial t} + \bm u_{0a} \cdot \nabla,
    \label{eq:convectop0}
\end{equation}

\noindent is related to the guiding-center fluid velocity $\bm u_{0a}$

\begin{equation}
    \bm u_{0a} = \bm U_{0a} + \frac{T_{\parallel a}+T_{\perp a}}{m_a}\frac{\bm b \times \nabla B}{\Omega_a B} +\frac{\bm b}{\Omega_a}\times \frac{d_{0 a} \bm U_{0 a}}{dt},
    \label{eq:guidvel}
\end{equation}

\noindent and it differs from the lower-convective derivative ${d_{0a}}/{dt}$ in \cref{eq:convdev0} by the addition of the last two terms in \cref{eq:guidvel}.
The vorticity $\nabla_\perp^2 \phi$ is related to the $\bm E \times \bm B$ drift by

\begin{equation}
    \frac{\bm b \cdot \nabla \times \bm v_E}{\Omega_a} = \frac{\nabla_\perp^2 \phi}{B \Omega_a} + O(\epsilon^3),
    \label{eq:vorticityapprox}
\end{equation}

\noindent and it appears in \cref{eq:continuity} due to the difference between $\overline N_a^{00}$ and $N_a^{00}$ [see \cref{eq:overlinenapj}].
To derive \cref{eq:continuity}, we use the low-$\beta$ limit expression for $\bm b \times \bm k \simeq (\bm b \times \nabla B)/B$ and neglect $u_{\parallel a}\bm b \cdot \nabla \times \bm b/\Omega_a$ as

\begin{equation}
    \frac{u_{\parallel a} \bm b \cdot \nabla \times \bm b}{\Omega_a} \sim \frac{T_e}{T_i}\beta \sim \epsilon^3,
    \label{eq:bcurlbapprox}
\end{equation}

\noindent therefore keeping up to $O(\epsilon^2)$ terms [namely the $\nabla_\perp^2 \phi$ term in \cref{eq:vorticityapprox}].

The parallel momentum equation is obtained by setting $(p,j)=(1,0)$ in \cref{eq:finalDKE}, yielding

\begin{equation}
    \begin{split}
        m_a\frac{d_a^0 u_{\parallel a}}{dt} &=\frac{m_a v_{th\parallel a}}{\sqrt 2}\sum_b C_{ab}^{10}-\frac{m_a \nabla_\perp^2 \phi}{\Omega_a B}\frac{d_0 u_{\parallel a}}{dt} - \frac{m_a}{\sqrt 2N_a}\nabla \cdot\left(\bm u_a^1 N_a v_{th\parallel a} \right) \\
        &+m_a ||\mathcal{A}||_a^{*00}+\left(1+\frac{\nabla_\perp^2 \phi}{\Omega_a B}\right)\left(q_a E_\parallel - T_{\perp a} \frac{\nabla_\parallel B}{B}+m_a \bm v_E \cdot \frac{d_{0 a} \bm b}{dt}\right),
    \end{split}
    \label{eq:parallelvelgc}
\end{equation}

with 

\begin{equation}
    \begin{split}
        \bm u_a^1 &= \frac{\bm U_{p a}^{th}}{\sqrt{2}}+\frac{\sqrt 2}{m_a}\frac{\bm b \times \nabla B}{\Omega_a B}\frac{Q_{\parallel a}+Q_{\perp a}}{N_a v_{th\parallel a}}+v_{th\parallel a}\frac{\bm b}{2}\left(1+\frac{\nabla_\perp^2 \phi}{\Omega_a B}\right).
    \end{split}
\end{equation}

\noindent The expression for $C_{ab}^{10}$ is given in Appendix \ref{app:cabmoments}, as well as all the $C_{ab}^{pj}$ coefficients relevant for the present fluid model.
The left-hand side of \cref{eq:parallelvelgc} describes the convection of $u_{\parallel a}$, while the first term in the right-hand side is related to pressure and heat flux gradients, the second term to resistivity (collisional effects), the third term consists of high-order terms kept to ensure phase-space conservation properties, and the last term is the parallel fluid acceleration, namely due to parallel electric fields, mirror force, and inertia.

The parallel and perpendicular temperature equations are obtained by setting $(p,j)=(2,0)$ and $(0,1)$ respectively in \cref{eq:finalDKE}. This yields for the parallel temperature

\begin{equation}
    \begin{split}
        \frac{N_a}{\sqrt{2}}\frac{d_a^0 T_{\parallel a}}{dt} &=
        \sqrt{2} Q_{\perp a} \frac{\nabla_\parallel B}{B}- \frac{N_a \nabla_\perp^2 \phi}{\sqrt{2}\Omega_a B}\frac{d_{0a} T_{\parallel a}}{dt}-2\frac{N_a T_{\parallel a}}{v_{th\parallel a}}\bm u_a^{1} \cdot \nabla u_{\parallel a}\\
        &-\nabla \cdot (N_a T_{\parallel a} \bm u_a^{2\parallel})+ N_a T_{\parallel a}\frac{\bm E}{B}\cdot \frac{\bm b \times \nabla B}{B}\left(1+\frac{\nabla_\perp^2 \phi}{\Omega_a B}\right)\\
        &+\sum_b C_{ab}^{20}N T_{\parallel a} +\frac{2 N_a T_{\parallel a}}{v_{th\parallel a}}||\mathcal{A}||_a^{*10},
    \end{split}
\label{eq:paralleltempc}
\end{equation}

\noindent where

\begin{equation}
    \begin{split}
        \bm u_a^{2\parallel}&= \frac{Q_\parallel a}{2 N_a T_{\parallel a}}\frac{\bm U_{pa}^{th}}{v_{th\parallel a}}+\frac{\sqrt{2}T_{\parallel a}}{m_a}\frac{\bm b \times \nabla B}{\Omega_a B}+\frac{\bm b}{2}\frac{Q_{\parallel a} }{N_a T_{\parallel a} }\left(1+\frac{\nabla_\perp^2 \phi}{\Omega_a B}\right),
    \end{split}
\end{equation}

\noindent and for the perpendicular temperature

\begin{equation}
    \begin{split}
        &N_a\frac{d_a^0 }{dt}\left(\frac{T_{\perp a}}{B}\right)+\frac{N_a\nabla_\perp^2 \phi}{\Omega_a B}\frac{d_{0 a} }{dt}\left(\frac{T_{\perp a}}{B}\right) =\nabla \cdot \left(\frac{N_a T_{\perp a}}{B} \bm u_a^{2\perp}\right)-\frac{ N_a T_{\perp a}}{B}\sum_b  C_{ab}^{01},
    \end{split}
    \label{eq:perptempc}
\end{equation}

\noindent with

\begin{equation}
    \begin{split}
        \bm u_a^{2\perp}&=-\frac{Q_{\perp a}}{N_a T_{\perp a}}\frac{\bm U_{pa}^{th}}{v_{th\parallel a}}-\frac{T_{\perp a}}{m_a}\frac{\bm b \times \nabla B}{\Omega_a B}.
    \end{split}
\end{equation}

The equations for the evolution of the parallel $Q_{\parallel a}$ and perpendicular $Q_{\perp a}$ heat fluxes are obtained by setting $(p,j)=(3,0)$ and $(1,1)$ respectively in \cref{eq:finalDKE}, yielding

\begin{equation}
    \begin{split}
        \frac{d_a^0 Q_{\parallel a}}{dt} &=-  \frac{d_{0 a}}{dt}\left(Q_{\parallel a}\frac{\nabla_\perp^2 \phi}{\Omega_a B}\right)+N_a T_{\parallel a} \sqrt{3} v_{th\parallel a} \sum_b C_{ab}^{30} \\
        &-Q_{\parallel a}\nabla \cdot \bm u_a^0- \frac{Q_{\parallel a} \nabla_\perp^2 \phi}{\Omega_a B}\nabla \cdot \bm U_{0 a} - 3 \nabla \cdot (\bm u_{k a} Q_{\parallel a})\\
        &-\frac{3}{\sqrt{2}}\left(1+\frac{\nabla_\perp^2 \phi}{\Omega_a B}\right)\frac{\bm E \cdot \bm b \times \nabla B}{B^2}Q_{\parallel a}+3 \sqrt{2}N_a T_{\parallel a} ||\mathcal{A}||_a^{*20}\\
        &-3 \sqrt{2}N_a T_{\parallel a} \bm u_a^{2\parallel} \cdot \nabla u_{\parallel a} - 3 \sqrt{2}N_a v_{th\parallel a} \bm u_a^1 \cdot \nabla T_{\parallel a},
    \end{split}
    \label{eq:qpar}
\end{equation}

\noindent and

\begin{equation}
    \begin{split}
        \frac{d_a^0}{dt}\left(\frac{Q_\perp a}{B}\right)&=- \frac{d_{0 a}}{dt}\left(\frac{Q_{\perp a}}{B}\frac{\nabla_\perp^2 \phi}{\Omega_a B}\right)-\frac{N_a v_{th\parallel a}}{\sqrt{2}}(\bm u_a^1 \cdot \nabla) \frac{T_{\perp a}}{B}\\
        &+\frac{N_a T_{\perp a}}{B}(\bm u_a^{2\perp} \cdot \nabla) u_{\parallel a}-\left(\frac{Q_{\perp a}}{B}\right)\left(\nabla \cdot \bm u_a^0 + \frac{\nabla_\perp^2 \phi}{\Omega B}\nabla \cdot \bm U_{0 a}\right)\\
        &-(\bm U_{k a} + 2 \bm U_{\nabla B})\cdot \nabla \left(\frac{Q_{\perp a}}{B}\right)-\frac{\sum_b C_{ab}^{11}}{\sqrt{2}}\frac{v_{th\parallel a}N_a T_{\perp a}}{B}\\
        &+\left(\frac{N_a T_{\perp a}^2}{m_a}\frac{\nabla_\parallel B}{B^2}+\frac{Q_{\perp a}}{B}\bm E \cdot \frac{\bm b \times \nabla B}{B^2}\right)\left(1+\frac{\nabla_\perp^2 \phi}{\Omega_a B}\right).
    \end{split}
    \label{eq:qperp}
\end{equation}

\noindent {In \cref{eq:qpar,eq:qperp} we neglected the higher-order moments with respect to $N^{30}$ and $N^{11}$, an approximation that we will scrutinize in the next section.}
Equations (\ref{eq:continuity})-(\ref{eq:qperp}) constitute a closed set of six coupled non-linear partial differential equations for both the hydrodynamical variables $n_a, u_{\parallel a}, T_{\parallel a}, T_{\perp a}$, and the kinetic variables $Q_{\parallel a}$ and $Q_{\perp a}$.

With respect to previous delta-F \citep{Dorland1993,Brizard1994} and full-F gyrofluid models \citep{Madsen2013}, our fluid model, Eqs. (\ref{eq:continuity}-\ref{eq:qperp}), while neglecting $k_{\perp} \rho_i \sim 1$ effects, {includes the velocity contributions from} the $B_{\parallel}^*$ denominator in the equations of motion (\ref{eq:GC1}) and (\ref{eq:GC2}) and includes the effects of full Coulomb collisions up to order $\epsilon_\nu \epsilon$.
{Also, due to the choice of basis functions with shifted velocity arguments $H_p(s_{\parallel a})$ instead of $H_p(v_\parallel/v_{tha})$, we obtain a set of equations that can efficiently describe both weak flow ($u_{\parallel a} \ll v_{th a})$ and strong flow ($u_{\parallel a} \sim v_{th a}$) regimes}.

\subsection{High Collisionality Regime}
\label{sec:highcoll}

We now consider the high-collisionality regime, where the characteristic fluctuation frequency of the hydrodynamical variables $\omega{}$

\begin{equation}
\begin{split}
	&\omega \sim v_{tha} |\nabla_\parallel \ln N_a| \sim v_{tha} |\nabla_\parallel \ln T_{\parallel a}|\sim v_{tha} |\nabla_\parallel \ln T_{\perp a}| \sim |\nabla_\parallel \ u_{\parallel a}| \sim v_{th a}/ L_{\parallel a},
\end{split}
\end{equation}

\noindent is much smaller than the collision frequency $\nu_a \simeq \nu_{aa}$, that is

\begin{equation}
    \delta_a \sim \frac{\omega}{\nu_a} \sim \frac{\lambda_{mfp a}}{L_{\parallel a}} \ll 1,
\label{eq:smallmfp}
\end{equation}

\noindent where the mean free path $\lambda_{mfp a}$ in \cref{eq:smallmfp} is defined as

\begin{equation}
    \lambda_{mfp a} = v_{th a}/\nu_{aa}.
\end{equation}

\noindent Equation (\ref{eq:smallmfp}) describes the so-called linear transport regime \citep{Balescu1988}.
In this case, the distribution function can be expanded around a Maxwell-Boltzmann equilibrium, according to the Chapman-Enskog asymptotic closure scheme \citep{Chapman1962} and, to first order in $\delta_a$, we have

\begin{equation}
    \lb F_a \rb \simeq F_{Ma}\left[1+\delta_a f_{1 a}(v_{\parallel},\mu,\bm R, t)\right].
\label{eq:chapenskexpansion}
\end{equation}

\noindent According to \cref{eq:chapenskexpansion}, all moments $N_a^{pj}$ in the Hermite-Laguerre expansion \cref{eq:gyrof} with $(p,j)\not=(0,0)$ are order $\delta_a$. {Since $Q_{\parallel a}$ and $Q_{\perp a}$ are determined at first order in $\delta_a$ only by the moments $(p,j)=(0,0),(3,0),(1,1)$, the truncation of Sec. (\ref{sec:fluideqs}), i.e., neglecting $(p,j)\not=(0,0),(3,0),(1,1)$ is justified. For a more detailed discussion {on this topic} see \citet{Balescu1988}.
Moreover, in the linear regime, a relationship between the hydrodynamical and kinetic variables can be obtained along the lines of the semi-collisional closure. This allows us to express $Q_{\parallel a}$ and $Q_{\perp a}$ as a function of $N_a, u_{\parallel a}, T_{\parallel a}$ and $T_{\perp a}$, therefore reducing the number of equations.
We now derive this functional relationship.

We consider Eqs. (\ref{eq:qpar})-(\ref{eq:qperp}) in the linear regime, and neglect the polarization terms that are proportional to $\nabla_\perp^2 \phi/(\Omega_a B)$.
{This yields $\sqrt{{3}/{2}}{\sum_b C_{ab}^{30}}/{v_{th\parallel a}} \simeq R_{\parallel a}$ and ${\sum_b C_{ab}^{11}}/(\sqrt{{2}}{v_{th\parallel a}}) \simeq R_{\perp a}$, with $R_{\parallel a}$ and $R_{\perp a}$ given by}

\begin{equation}
    R_{\parallel a} = \frac{\nabla_{\parallel} T_{\parallel a}}{T_{\parallel a}} +u_{\parallel a} \frac{\bm b \times \nabla B}{\Omega_a B}\cdot \left(\frac{\nabla u_{\parallel a}}{u_{\parallel a}}+\frac{\nabla T_{\parallel a}}{T_{\parallel a}}\right),
    \label{eq:collimit1}
\end{equation}
\begin{equation}
\begin{split}
    R_{\perp a} &= \frac{T_{\perp a}}{T_{\parallel a}}\frac{\nabla_{\parallel} B}{B} - \frac{1}{2\sqrt{2}}\nabla_{\parallel} \ln \frac{T_{\perp a}}{B}-u_{\parallel a} \frac{\bm b \times \nabla B}{\Omega_a B}\cdot \left(\frac{T_{\perp a}}{T_{\parallel a}}\frac{\nabla u_{\parallel a}}{u_{\parallel a}}+\nabla \ln \frac{T_{\perp a}}{B}\right),
\end{split}
\label{eq:collimit2}
\end{equation}

\noindent since $d_a^0/dt \sim d_{0 a}/dt \sim \omega{}$ and $(d^0 Q_{\parallel, \perp }/dt)/Q_{\parallel,\perp a} \sim \delta_a^2 \nu_a$.
We compute the guiding-center moments of the collision operator $C_{ab}^{30}$ and $C_{ab}^{11}$ by truncating the series for the like-species collision operator in \cref{eq:caapjexact} at $(l,k,n,q)=(2,1,2,1)$.
The resulting $C_{ab}^{pj}$ coefficients are presented in Appendix \ref{app:cabmoments}.

With the expression of $C_{ab}^{30}$ and $C_{ab}^{11}$, we can solve for $Q_{\parallel a}$ and $Q_{\perp a}$. In the regime $(T_{\parallel a}-T_{\perp a})/T_a \sim \delta$, at lowest order, we obtain for the electron species

\begin{equation}
    \frac{Q_{\parallel e}}{N_e T_{ e} v_{th e}} =-0.362 \frac{u_{\parallel e}-u_{\parallel i}}{v_{th e}}-10.6 \lambda_{mfpe}\frac{\nabla_\parallel T_e}{T_e},
    \label{eq:qpare}
\end{equation}

\noindent and

\begin{equation}
    \frac{Q_{\perp e}}{N_e T_{ e} v_{th e}} =-0.119 \frac{u_{\parallel e}-u_{\parallel i}}{v_{th e}}-3.02 \lambda_{mfpe}\frac{\nabla_\parallel T_e}{T_e},
\label{eq:qperpe}
\end{equation}

\noindent Analogous expressions are obtained for the ion species.

Equations (\ref{eq:continuity}), (\ref{eq:parallelvelgc}), (\ref{eq:paralleltempc}), and (\ref{eq:perptempc}), with $Q_{\parallel a}$ and $Q_{\perp a}$ given by \cref{eq:qperpe,eq:qpare} are valid in the high-collisionality regime, and can be compared with the drift-reduced Braginskii equations in \citet{Zeiler1997}. We first rewrite the continuity equation, \cref{eq:continuity}, in the form
\begin{equation}
    \frac{\partial N_e}{\partial t} + \nabla \cdot \left[N_e\left(\bm v_E + u_{\parallel e} \bm b+  \frac{T_{\parallel e}+T_{\perp e}}{m_e}\frac{\bm b \times \nabla B}{\Omega_e B} \right)\right]
    =0,
\end{equation}
\noindent where we expand the convective derivative $d^0{a}/dt$ using \cref{eq:convectop0} and \cref{eq:guidvel}, and neglect polarization terms proportional to the electron mass $m_e$. By noting that the diamagnetic drift $v_{de}$ can be written as 
\begin{equation}
    \bm v_{de} = \frac{1}{e N_e}\nabla \times \frac{p_e \bm b}{B}-2\frac{T_e}{m_e}\frac{\bm b \times \nabla B}{\Omega_e B},
\end{equation}
\noindent and by considering the isotropic regime $T_{\parallel e} \sim T_{\perp e} \sim T_{e}$, we obtain
\begin{equation}
    \frac{\partial N_e}{\partial t} + \nabla \cdot \left[N_e\left(\bm v_E + u_{\parallel e} \bm b+  \bm v_{de} \right)\right]
    =0,
\end{equation}
\noindent which corresponds to the continuity equation in the drift-reduced Braginskii model in \citet{Zeiler1997}. In {that model}, the polarization equation is obtained by subtracting both electron and ion continuity equations, using Poisson's equation $n_e \simeq n_i$ with $n_e$ and $n_i$ the electron and ion particle densities respectively, and neglecting the electron to ion mass ratio. Applying the same procedure to the present fluid model, we obtain

\begin{equation}
\begin{split}
    0&=\nabla \cdot \left(\frac{\nabla_\perp^2 \phi N_i u_{\parallel i} \bm b}{\Omega_i B}\right)-\nabla \cdot \left[\frac{\bm v_E}{2 m_i}\nabla_\perp^2\left(\frac{N_i T_{\perp i}}{\Omega_i^2}\right)\right]-\frac{1}{2m_i}\frac{\partial}{\partial t}\nabla_\perp^2 \left(\frac{N_i T_{\perp i}}{\Omega_i^2}\right)\\
    &+\nabla \cdot \left(\frac{N_i}{\Omega_i}\bm b \times \frac{d_{0i}\bm U_{0i}}{dt}\right)+\nabla \cdot \left[\bm b\left(N_i u_{\parallel i} - N_e u_{\parallel e}\right)\right]\\
    &+\nabla \cdot \left[\left({N_i T_{\parallel i}+N_eT_{\parallel e}+N_i T_{\perp i}+N_eT_{\perp e}}\right)\frac{\bm b \times \nabla B}{e B^2}\right].
\end{split}
\label{eq:gcpolarizationeq}
\end{equation}

In \cref{eq:gcpolarizationeq}, the first three terms, which are not present in the drift-reduced Braginskii model, correspond to the difference between ion guiding-center density $N_i$ and particle density $n_i$, proportional to both $\nabla_\perp^2 \phi$ and $\nabla_\perp^2 P_i$.
The parallel momentum and temperature equations, \cref{eq:parallelvelgc} and \cref{eq:paralleltempc}, with respect to \citep{Zeiler1997}, contain the higher-order term $\mathcal{A}$ that ensures phase-space conservation, mirror force terms proportional to $(\nabla_\parallel B)/B$, and polarization terms proportional to $\nabla_\perp^2 \phi/(\Omega_a B)$ due to the difference between guiding-center and particle fluid quantities.
{This set of fluid equations constitute an improvement over the drift-reduced Braginskii model. With respect to the original Braginskii equations \citep{Braginskii1965}, they include the non-linear terms that arise when retaining full Coulomb collisions, and the effect of ion-electron collisions.}

\section{Conclusion}

In the present work, a full-F drift-kinetic model is developed, suitable to describe the plasma dynamics in the SOL region of tokamak devices at arbitrary collisionality. Taking advantage of the separation between the turbulent and gyromotion scales, a gyroaveraged Lagrangian and its corresponding equations of motion are obtained. This is the starting point to deduce a drift-kinetic Boltzmann equation with full Coulomb collisions for the gyroaveraged distribution function.

The gyroaveraged distribution function is then expanded into an Hermite-Laguerre basis, and the coefficients of the expansion are related to the lowest-order gyrofluid moments. The fluid moment expansion of the Coulomb operator described in \citet{Ji2009} is reviewed, and its respective particle moments are written in terms of coefficients of the Hermite-Laguerre expansion, relating both expansions. This allows us to express analytically the moments of the collision operator in terms of guiding-center moments.
A moment hierarchy that describes the evolution of the guiding-center moments is derived, together with a Poisson's equation accurate up to $\epsilon^2$. These are then used to derive a fluid model in the high-collisionality limit.

The drift-kinetic model derived herein can be considered a starting point for the development of a gyrokinetic Boltzmann equation suitable for the SOL region (e.g. \citet{Qin2007, Hahm2009}). Indeed, using a similar approach, a gyrokinetic moment hierarchy may be derived, allowing for the use of perpendicular wave numbers satisfying $k_\perp \rho_s \sim 1$. {For a recent Hermite-Laguerre formulation of the non-linear delta-F gyrokinetic equation see \citet{Mandell2017}}.

\section{Acknowledgements}
This work has been carried out within the framework of the EUROfusion Consortium and has received funding from the Euratom research and training programme 2014-2018 under grant agreement No 633053, and from Portuguese FCT - Fundação para a Ciência e Tecnologia, under grant PD/BD/105979/2014.
N.F.L. was partially funded by US Department of Energy Grant no. DE-FG02-91ER54109.
The views and opinions expressed herein do not necessarily reflect those of the European Commission.

\appendix
\addtocontents{toc}{\protect\setcounter{tocdepth}{0}}

\section{Basis Transformation}
\label{app:tlkpj}

In the present Appendix, we derive the expressions for the coefficients $T_{alk}^{pj}$ appearing in \cref{eq:tlkpj}. These coefficients allows us to express up to order $\epsilon \epsilon_\nu$ the relation between fluid $\bm M_{a}^{lk}$ and guiding-center $N_{a}^{lk}$ moments via \cref{eq:CoulDKmom}.
As a first step, we define a transformation similar to \cref{eq:tlkpj} but with isotropic temperatures between both {bases}

\be
    \begin{split}
        c_a^l P_l(\xi_a)L_k^{l+1/2}(c_a^2)=&\sum_{p=0}^{l+2k}\sum_{j=0}^{k+\floor{l/2}}\overline T_{lka}^{pj} H_p\left(\frac{v_\parallel-u_{\parallel a}}{v_{th a}}\right) L_j\left(\frac{v^{'2}_{\perp}}{v_{th a}^2}\right),
    \end{split}
    \label{eq:deftbarlkpj}
\ee

\noindent with the inverse transformation

\be
    \begin{split}
        H_p\left(\frac{v_\parallel-u_{\parallel a}}{v_{th a}}\right)L_j\left(\frac{v^{'2}_{\perp}}{v_{th a}^2}\right) =& \sum_{l=0}^{p+2j}\sum_{k=0}^{j+\floor{p/2}}\l(\overline T^{-1}\r)_{pj a}^{lk} \\
        &\times c_a^l P_l(\xi_a)L_k^{l+1/2}(c_a^2),
    \end{split}
\ee

The relation between the coefficients $\l(\overline T^{-1}\r)_{pj}^{lk}$ and $\overline T_{lk}^{pj}$ is given by

\be
    \l(\overline T^{-1}\r)_{pj}^{lk}=\frac{\sqrt{\pi}2^p p!(l+1/2)k!}{(k+l+1/2)!}\overline T_{lk}^{pj}.
\ee

By integrating both sides of \cref{eq:deftbarlkpj} over the whole velocity space, the expression for $\overline{T}_{lk}^{pj}$ is obtained

\begin{equation}
    \begin{split}
        \overline T_{lk}^{pj}&=\sum_{q=0}^{\floor{l/2}}\sum_{v=0}^{\floor{p/2}}\sum_{i=0}^{k}\sum_{r=0}^{q}\sum_{s=0}^{\text{min}(j,i)}\sum_{m=0}^{k-i}\frac{(-1)^{q+i+j+v+m}}{2^{\frac{3l+p}{2}+m+v-r}}\\
        &\times\binom{l}{q}\binom{2(l-q)}{l}\binom{q}{r}\binom{r}{j-s}\binom{r}{i-s}\binom{s+r}{s}{r!}{}\\
        &\times\frac{(k-i+l-1/2)!(l+p+2(m-r-v)-1)!!}{(p-2v)!(k-i-m)!(l+m-1/2)!v!m!}.
    \end{split}
\end{equation}

We then integrate both sides of \cref{eq:tlkpj} with weights $H_l(s_{\parallel a})L_j(s_{\perp a}^2)$, with the argument transformation

\begin{equation}
\begin{split}
        H_p(s_{\parallel a})=\left(\frac{T_a}{T_{\parallel a}}\right)^{p/2}\sum_{k=0}^{\floor{p/2}}&\frac{p!}{k!(p-2k)!}\left(1-\frac{T_{\parallel a}}{T_a}\right)^kH_{p-2k}\left(\frac{v_\parallel-u_{\parallel a}}{v_{th a}}\right),
\end{split}
\end{equation}

\noindent and

\begin{equation}
\begin{split}
    L_j(s_{\perp a}^2)=\sum_{k=0}^{j}&\binom{j}{j-k}\left(\frac{T_a}{T_{\perp a}}\right)^k \left(1-\frac{T_a}{T_{\perp a}}\right)^{j-k}L_{k}\left(\frac{v_{\perp}^{'2}}{v_{th a}^2}\right),
\end{split}
\end{equation}

\noindent to find the relation between the isotropic and anisotropic temperature coefficients

\be
    \begin{split}
        T_{alk}^{pj}=&\sum_{pp=0}^{l+2k}\sum_{jj=0}^{k+\floor{l}{2}}\sum_{z=0}^{jj}\sum_{d=0}^{\floor{pp/2}}\binom{jj}{jj-z}\frac{pp!\delta_{z,j}\delta_{p,pp-2d}}{d!(pp-2d)!}\\
        &\times\l(\frac{T_{\parallel a}}{T_a}\r)^{p/2}\l(\frac{T_{\perp a}}{T_a}\r)^{z}\l(1-\frac{T_a}{T_{\parallel a}}\r)^{d}\l(1-\frac{T_{\perp a}}{T_a}\r)^{jj-z}{\overline{T}_{lk}^{pp jj}},
    \end{split}
    \label{eq:tlkpjexact}
\ee

\be
    \begin{split}
        \left(T^{-1}_a\right)_{pj }^{lk}=&\sum_{z=0}^{j}\sum_{d=0}^{\floor{p/2}}\sum_{ll=0}^{p-2d+2z}\sum_{kk=0}^{z-d+\floor{p}{2}}\binom{j}{j-z}\frac{p!\delta_{l,ll}\delta_{k,kk}}{d!(p-2d)!}\\
        &\times\l(\frac{T_{ a}}{T_{\parallel a}}\r)^{p/2}\l(\frac{T_{a}}{T_{\perp a}}\r)^{z}\l(1-\frac{T_{\parallel a}}{T_{ a}}\r)^{d}\l(1-\frac{T_{ a}}{T_{\perp a}}\r)^{j-z}\left(\overline{T^{-1}}\right)_{p-2d z}^{ll kk}.
    \end{split}
    \label{eq:tlkpjminus1exact}
\ee

\section{Guiding-Center Moments of $\mathcal{A}$}
\label{app:collcoef}

In \cref{eq:GC2}, the term $\mathcal{A}$ that ensures phase-space conservation properties for the particle equations of motion is introduced. Here, we present the analytic expressions for its guiding-center moments $|| \mathcal A||_a^{*pj}$ appearing in \cref{eq:finalDKE4}. These are given by

\begin{equation}
\begin{split}
    || \mathcal{A}||_a^{*pj} &= \frac{1}{N_a \Omega_a} \sum_{l,k}\left(A_{1 a}\mathcal{V}_{lk}^{3pj}+A_{2 a} \mathcal{V}_{lk}^{2pj}+A_{3 a} \mathcal{V}_{lk}^{1pj}\right.\\
    &\left.+A_{4 a} \mathcal{V}_{lk}^{1p'j'}\mathcal{M}_{p'j'}^{pj}+A_{5 a} \mathcal{M}_{lk}^{pj}+A_{6 a} \delta_{pl}\delta_{jk}\right)N_a^{lk},
\end{split}
\label{eq:mathavv}
\end{equation}

\noindent with the phase-mixing term

\be
    \begin{split}
        \mathcal{V}_{lk}^{3pj}=&\l[\sqrt{(p+3)(p+2)(p+1)}\delta_{p+3,l}+3\sqrt{(p+1)^3}{\delta_{p+1 ,l}}\right.\\
        &\left.+ 3\sqrt{p^3}{\delta_{p-1 ,l}}+\sqrt{p(p-1)(p-2)}\delta_{p-3,l} \r]\frac{\delta_{j,k}}{\sqrt{8}},
    \end{split}
    \label{eq:vv3pjlk}
\ee

\noindent and the coefficients $A_i$

\begin{align}
        A_{1 a}&= v_{th\parallel a}^3 \bm k_\perp \cdot \nabla \times \bm b,\\
        A_{2 a} &= v_{th\parallel a}^2 [\bm k_\perp \cdot (u_{\parallel a} \nabla \times \bm b + \nabla \times \bm v_E) +  \nabla \times \bm b \cdot \bm A],\\
        A_{3 a} &=v_{th\parallel a} (u_{\parallel a} \nabla \times \bm b + \nabla \times \bm v_E) \cdot \bm A + v_{th\parallel a}^2 \nabla \times \bm b \cdot \bm C,\\
        A_{4 a} &= v_{th\parallel a} \frac{T_\perp}{m_a B}\nabla_\perp B \cdot \nabla \times \bm b,\\
        A_{5 a} &= \frac{T_{\perp a}}{m_a B}\nabla_\perp B \cdot (u_{\parallel a} \nabla \times \bm b + \nabla \times \bm v_E),\\
        A_{6 a} &= (v_{th\parallel a} u_{\parallel a} \nabla \times \bm b + \nabla \times \bm v_E) \cdot \bm C
\end{align}

\noindent with

\begin{align}
        \bm A &= \left[ \frac{\partial \bm b}{\partial t} + (\bm b \cdot \nabla) \bm v_E + (\bm v_E \cdot \nabla) \bm b +2 u_{\parallel a} v_{th\parallel a} \bm k\right]_{\perp},\\
        \bm C &= \frac{1}{v_{th\parallel a}}\left[\frac{\partial \bm v_E}{\partial t}+(\bm v_E \cdot \nabla)\bm v_E + u_{\parallel a}^2 \bm k\right]_{\perp}.
\end{align}

\section{Poisson's Equation with Collisional Effects}
\label{app:poisson}

To include $\epsilon_\nu$ effects in Poisson's equation, we retain the $l=1$ Bessel term in \cref{eq:poissonpresqueexact}, yielding

\begin{equation}
\begin{split}
    \epsilon_0 \nabla \cdot \bm E&=\sum_a q_a\left[N_{a}\left(1+\frac{\bm b \cdot \nabla \times \bm b}{\Omega} u_{\parallel a} +\frac{\bm b \cdot \nabla \times \bm v_E}{\Omega}\right) \right.\\
    &\left.+\frac{1}{2m_a}\nabla_\perp^2 \left(\frac{P_{\perp a}}{\Omega_a^2}\right)+2\pi \int \Gamma_1 [C_{1 a}e^{i \alpha}+C_{-1 a}e^{-i \alpha}] \frac{B_\parallel^*}{m_a} dv_\parallel d\mu\right].
\end{split}
    \label{eq:poissonfin1}
\end{equation}

\noindent The collisional terms $C_{\pm 1 a} = \sum_b C_{\pm 1 ab}$ (for collisions between species $a$ and $b$) can be cast in terms of gyrofluid moments $N_a^{lk}$. For like-species collisions, we use \cref{eq:ccjiheld} to express the collision operator $C_{aa0}$ in \cref{eq:JiCab} in terms of fluid moments, together with the property \citep{Ji2006}

\begin{equation}
    \bm P^{l}(\bm v)\cdot \bm T^{lk} = \bm v^l \cdot \bm T^{lk},
\end{equation}

\noindent which holds for any totally symmetric and traceless tensor $\bm T^{lk}$.
This yields the following form for the lowest-order collision operator \cref{eq:ccjiheld}

\begin{equation}
    \begin{split}
    c_0\l(f_a^{lkm},f_a^{nqr}\r)&=f_{aM}\mathcal{N}_a^{lk}\mathcal{N}_a^{nq}\sum_{u=0}^{\text{min}(2,l,n)}\nu_{*aau}^{lm,nr}(c^2)\\
    &\times\sum_{i=0}^{\text{min}(l,n)-u}d_i^{l-u,n-u}\frac{\bm c_a^{l+n-2(i+u)}}{c_a^{l+n-2(i+u)}}\cdot \overline{\bm P^{l}(\bm b)\cdot^{i+u}\bm P^{n}(\bm b)},
    \end{split}
    \label{eq:ccjiheld1}
\end{equation}

\noindent as $\overline{\bm T}$ means the traceless symmetrization of $\bm T$.
The shifted velocity vector $\bm c_a = (\bm v - \bm u_a)/v_{tha}$, under the transformation of \cref{eq:GCcoordinates}, with the lowest-order fluid velocity $\bm u_a \simeq u_{\parallel a} \bm b + \bm v_E$, can be written as

\begin{equation}
    \bm c_a = c_{\parallel a} \bm b + \frac{c_{\perp a}}{2 }\left(e^{i \theta}\bm E_1+e^{-i\theta}\bm E_2\right),
    \label{eq:shiftedvelca}
\end{equation}

\noindent where $\bm E_{1,2}=\bm e_2 \pm i \bm e_1$. Using the multinomial theorem

\begin{equation}
    \left(\sum_{i=1}^m x_i\right)^k = 
    \sum_{a_i\ge0}\frac{k!\Pi_{i=1}^m x_i^{a_i}}{\Pi_{i=1}^m a_i!},
\end{equation}

\noindent subject to the constraint $\sum_{i=1}^m a_i = k$, we obtain

\begin{equation}
    \bm c_a^k=\sum_{a_1+a_2+a_3=k}\frac{k!(c_{\parallel a})^{a_1}}{a_1!a_2!a_3!} \left(\frac{c_{\perp a}}{2}\right)^{a_2+a_3}e^{i\theta(a_2-a_3)}\bm b^{a_{1}} \bm E_1^{a_2} \bm E_2^{a_3}.
    \label{eq:vtothek}
\end{equation}

We use \cref{eq:vtothek} to explicit the dependence of $\bm c_a$ on $\theta$ in \cref{eq:ccjiheld1}. The Fourier components $C_{1a}$ and $C_{-1a}$ correspond to the case $a_2=a_3\pm1$ in the sum in \cref{eq:vtothek} above, yielding

\begin{equation}
\begin{split}
    C_{\pm 1 a a}=&\sum_{l,k,m}\sum_{n,q,r}\frac{L_{km}^lL_{qr}^n}{\sqrt{\sigma_k^l \sigma_q^n}}f_{aM}\mathcal{N}_a^{lk}\mathcal{N}_a^{nq}\sum_{u=0}^{\text{min}(2,l,n)}\nu_{*aau}^{lm,nr}(c_a^2)\sum_{i=0}^{\text{min}(l,n)-u}d_i^{l-u,n-u}\\
    &\times\sum_{\mathclap{\substack{a_1+a_2+a_3=\\l+n-2(i+u)}}
    }\frac{(l+n-2(i+u))!}{a_1!a_2!a_3!}
    (c_{\parallel a})^{a_1}\left(\frac{c_{\perp a}}{2}\right)^{a_2+a_3}\delta_{a_2,a_3\pm1}\\
    &\times\frac{\bm b^{a_1} \bm E_1^{a_2} \bm E_2^{a_3}}{(c_a)^{l+n-2(i+u)}}\cdot \overline{\bm P^{l}(\bm b)\cdot^{i+u}\bm P^{n}(\bm b)}.
\end{split}
    \label{eq:ccjiheld2}
\end{equation}

Assembling the velocity dependent terms of \cref{eq:ccjiheld2}, together with $J_1(k_\perp \rho_a) \simeq k_\perp v_{tha} c_{\perp a}/(2 \Omega_a)$, the velocity integration of the like-species operator in Poisson's \cref{eq:poissonfin1} is then

\begin{equation}
\begin{split}
    I_{aa}^{\pm}&=\int c_{\perp a} C_{\pm 1 aa} \frac{B_\parallel^*}{m_a} dv_\parallel d\mu=\sum_{lkm}\sum_{nqr}\sum_{u=0}^{\text{min}(2,l,n)}\sum_{i=0}^{\text{min}(l,n)-u}\sum_{{a_1+a_2+a_3=l+n-2(i+u)}}\\
    &\times\frac{L_{km}^lL_{qr}^n}{\sqrt{\sigma_k^l \sigma_q^n}}
    \mathcal{N}_a^{lk}\mathcal{N}_a^{nq}d_i^{l-u,n-u}\delta_{a_2,a_3\pm1}\frac{(l+n-2(i+u))!}{a_1!a_2!a_3!2^{a_2+a_3}}\\
    &\times{\bm b^{a_1} \bm E_1^{a_2} \bm E_2^{a_3}}{}\cdot \overline{\bm P^{l}(\bm b)\cdot^{i+u}\bm P^{n}(\bm b)}v_{tha}^3 I_{a\pm},
\end{split}
\end{equation}

\noindent where

\begin{equation}
    I_{a\pm} = \int f_{M a} c_{\perp a} (c_{\parallel a})^{a_1}\left({c_{\perp a}}{}\right)^{2 a_2\pm1}\frac{\nu_{*aau}^{lm,nr}(c_a^2)}{c_a^{l+n-2(i+u)}}\frac{B_\parallel^*}{m_a} dv_\parallel d\mu.
    \label{eq:poissonintcaa1}
\end{equation}

Converting to pitch angle coordinates $v_{\parallel} =v_{tha} c_{\parallel a}= v_{tha} \xi_a c_a$, $v_\perp^{'2} = v_{tha}^2 c_{\perp a}^2 =c_a^2(1-\xi_a^2)$, with the volume element $(B_\parallel^*/m_a)d\mu dv_\parallel = v_{tha}^3 c_a^2 dc_a d\xi_a$, the integral of \cref{eq:poissonintcaa1} can be performed analytically, yielding

\begin{equation}
\begin{split}
    I_{a\pm}&=\int_{-1}^1 \xi_a^{a_1}(1-\xi_a^2)^{a_3+\sigma_\pm}d\xi_a\int f_{M a}\nu_{*aau}^{lm,nr}c^{2(a_3+\sigma_\pm+i+u)+a_1-l-n}\frac{d^3 \bm c}{4\pi}\\
    &=\frac{\left((-1)^{a_1}+1\right) \Gamma \left(\frac{a_1+1}{2}\right) \Gamma (a_3+\sigma_\pm +1)}{8 \pi \Gamma \left(\frac{a_1}{2}+a_3+\sigma_\pm +\frac{3}{2}\right)}C_{*aau}^{a_1-l-n~a_3+\sigma_\pm+i+u,lm,nr},
\end{split}
\end{equation}

\noindent where $\sigma_\pm = (1\pm1)/2$ for $C_{\pm1aa}$ respectively.

For electron-ion collisions, we take the expression for $C_{ei} = C_{ei}^0 + C_{ei}^1$ given by Eqs. (\ref{eq:cei0}) and (\ref{eq:cei1}) respectively. By using the shifted velocity vector $\bm c_a$ of \cref{eq:shiftedvelca}, we can proceed as for the like-species operator, yielding the Fourier components $C_{\pm 1 ei} = C_{\pm 1 ei}^0+C_{\pm 1 ei}^1$, with

\begin{equation}
\begin{split}
    C_{\pm 1 ei}^0&=-\sum_{l,k} \frac{n_i L_{ei}}{8 \pi c_e^3}\frac{l(l+1) f_{eM} \mathcal{N}_e^{lk}}{\sqrt{\sigma_k^l}}\\
    &\times\sum_{a_1+a_2+a_3=l}\frac{l!(c_{\parallel e})^{a_1}}{a_1!a_2!a_3!}\left(\frac{c_{\perp e}}{2}\right)^{a_2+a_3}\bm b^{a_1}\bm E_1^{a_2} \bm E_2^{a_3} \cdot \bm P^l(\bm b) \delta_{a_2,a_3\pm1},
\end{split}
\end{equation}

\noindent and

\begin{equation}
    C_{\pm 1 ei}^1= f_{Me} \frac{n_i L_{ei}m_e}{8 \pi c_e^3 T_e}\frac{c_{\perp e}}{2}\bm u_{ei} \cdot \bm E_{1,2}.
\end{equation}

The velocity integration of $C_{\pm 1 ei}^0$ and $C_{\pm 1 ei}^1$ are given by

\begin{equation}
\begin{split}
    I_{ei}^{0\pm}&=\int c_{\perp e} C_{\pm 1 ei}^0 \frac{B_\parallel^*}{m_a} dv_\parallel d\mu \\
    &=-\sum_{l,k}\sum_{a_1+a_2+a_3=l} \delta_{a_2,a_3\pm1} \frac{\nu_{ei}n_e v_{the}^3 \mathcal{N}_e^{lk}}{16 \pi^{3/2}\sqrt{\sigma_k^l}2^{a_2+a_3}}\frac{l(l+1)l!}{a_1!a_2!a_3!}\\
    &\times\frac{\Gamma\left(\frac{a_1+a_3+\sigma_\pm}{2}\right)\Gamma\left(\frac{a_1+1}{2}\right)\Gamma\left(a_3+\sigma_\pm+1\right)}{\Gamma\left(\frac{3}{2}+\frac{a_1}{2}+a_3+\sigma_\pm\right)},
\end{split}
\end{equation}

\noindent and

\begin{equation}
\begin{split}
    I_{ei}^{1\pm}=\int c_{\perp e} C_{\pm 1 ei}^1 \frac{B_\parallel^*}{m_a} dv_\parallel d\mu = \frac{n_e \nu_{ei}}{6 \sqrt{\pi}} v_{the}^2 \bm u_{ei} \cdot \bm E_{1,2},
\end{split}
\end{equation}

\noindent respectively. Ion-electron collisions are neglected due to the smallness of the electron to ion mass ratio.
Poisson's equation including $\epsilon^2$ and $\epsilon_\nu \epsilon$ effects then reads

\begin{equation}
\begin{split}
    \epsilon_0 \nabla \cdot \bm E&=\sum_a q_a\left[N_{a}\left(1+\frac{\bm b \cdot \nabla \times \bm b}{\Omega_a} u_{\parallel a} +\frac{\bm b \cdot \nabla \times \bm v_E}{\Omega_a}\right)\right.\\
    &+\left.\frac{1}{2m_a}\nabla_\perp^2 \left(\frac{P_{\perp a}}{\Omega_a^2}\right)+\sum_b \frac{\pi v_{tha}}{\Omega_a} \int k_\perp\left(e^{i \alpha} I_{ab}^+ + e^{-i \alpha} I_{ab}^-\right)e^{i \bm k \cdot \bm x}d^3\bm k\right].
\end{split}
    \label{eq:poissonfin3}
\end{equation}

\section{Expressions for the Moments of the Collision Operator}
\label{app:cabmoments}

In the present Appendix, we present the expressions for the guiding-center moments of the collision operator relevant for the fluid model in \cref{sec:fluidmodel}.
The collision operator moments satisfy particle conservation
\be
    C_{ab}^{00}=0,
\ee

\noindent and momentum conservation {at lowest order}

\be
    C_{aa}^{10}=0,
\ee

\begin{equation}
    C_{ei}^{10}=-\frac{m_i}{m_e}\frac{v_{th\parallel i}}{v_{th \parallel e}}C_{ie}^{10}{+O({m_e/m_i})}.
\end{equation}

\noindent Both the like-species and electron-ion satisfy energy conservation exactly, {while the ion-electron operator satisfies \cref{eq:cabenergy} at zeroth order in $\delta_a$}

\begin{equation}
    T_{\parallel a}C_{ab}^{20}-\sqrt{2}T_{\perp a} C_{ab}^{01} = 0.
    \label{eq:cabenergy}
\end{equation}

{The remaining moments $C_{ab}^{pj}$, in the linear transport regime with $\Delta T_a/T_a = (T_{\parallel a} - T_{\perp a})T_a \sim N^{11}\sim N^{30} \sim (u_{\parallel e}-u_{\parallel i})/v_{the} \sim \delta_a$, for ion-electron collisions are given by}

\begin{align}
    C_{ie}^{10}&=-\frac{m_e}{m_i}\frac{v_{th \parallel i}}{v_{th \parallel e}}C_{ei}^{10},\\
    C_{ie}^{20}&=\sqrt{2}\nu_{ei}\frac{m_e}{m_i}\left(\frac{T_e-T_i}{T_i}\right)-\frac{2 \sqrt{2} \nu_{ei}}{3}\frac{m_e}{m_i}\frac{T_e}{T_i}\frac{\Delta T_i}{T_i},\\
    C_{ie}^{01}&=-2\nu_{ei}\frac{m_e}{m_i}\left(\frac{T_e-T_i}{T_i}\right)-\frac{2 \nu_{ei}}{3}\frac{m_e}{m_i}\frac{T_e}{T_i}\frac{\Delta T_i}{T_i},\\
    C_{ie}^{30}&=-\nu_{ei}\sqrt{\frac{3}{2}}\frac{m_e}{m_i}\frac{Q_{\parallel i}}{n T_i v_{thi}},\\
    C_{ie}^{11}&=3 \nu_{ei}\frac{m_e}{m_i}\frac{Q_{\perp i}}{n T_i v_{thi}},
\end{align}

\noindent {for electron-ion collisions}

{
\begin{align}
    C_{ei}^{10} &= -\frac{\sqrt{2}\nu_{ei}}{6 \pi^{3/2}}\frac{u_{\parallel e}-u_{\parallel i}}{v_{the}}+\frac{\sqrt{2}\nu_{ei}}{10 \pi^{3/2}}\frac{Q_{\parallel e}+2 Q_{\perp e}}{n T_e v_{the}},\\
    C_{ei}^{20}&=-\frac{2 \sqrt{2}\nu_{ei}}{15 \pi^{3/2}}\frac{\Delta T_e}{T_e},\\
    C_{ei}^{30}&= \frac{\sqrt{3}\nu_{ei}}{10 \pi^{3/2}}\frac{u_{\parallel e}-u_{\parallel i}}{v_{the}}-\frac{ \nu_{ei}}{70 \sqrt{3} \pi^{3/2}}\frac{31 Q_{\parallel e} - 2 Q_{\perp e}}{n T_e v_{the}},\\
    C_{ei}^{11}&= \frac{ \nu_{ei}}{5 \sqrt{2} \pi^{3/2}}\frac{u_{\parallel e}-u_{\parallel i}}{v_{the}}+\frac{ \nu_{ei}}{150 \sqrt{2} \pi^{3/2}}\frac{Q_{\parallel e}-94 Q_{\perp e}}{n T_e v_{the}},\\
\end{align}
}
\noindent {and for like-species collisions}
{
\begin{align}
    C_{aa}^{20}&=0,\\
    C_{aa}^{30}&=-\frac{2 \sqrt{2}}{125 \sqrt{3} \pi^{3/2}}\frac{\nu_{aa}}{n T_a v_{tha}}\left(19 Q_{\parallel a}-7 Q_{\perp a}\right),\\
    C_{aa}^{11}&=-\frac{2}{375 \pi^{3/2}}\frac{\nu_{aa}}{n T_a v_{tha}}\left(7 Q_{\parallel a}-121 Q_{\perp a}\right).
\end{align}
}
\bibliographystyle{jpp}
\bibliography{library}

\end{document}